\documentclass[twocolumn]{aastex631}

\usepackage[utf8]{inputenc}
\usepackage{graphicx}

\newcommand\g{\; {\rm g}}

\newcommand\Msun{\;{\rm M_\odot}}
\newcommand\kms{\; {\rm km}\;{\rm s}^{-1}}

\newcommand\yr{\; {\rm yr}}
\newcommand\Myr{\;{\rm Myr}}

\newcommand\pc{\;{\rm pc}}
\newcommand\kpc{\;{\rm kpc}}

\newcommand\Kel{\;{\rm K}}

\newcommand{\hide}[1]{}

\begin{document}

\title{Cloud properties in simulated galactic winds}

\author[0009-0008-7057-9683]        {Orlando Warren}
\affil{Department of Physics \& Astronomy and Pitt PACC, University of Pittsburgh, 100 Allen Hall, 3941 O'Hara Street, Pittsburgh, PA 15260, USA}

\author[0000-0001-9735-7484]{Evan E. Schneider}
\affil{Department of Physics \& Astronomy and Pitt PACC, University of Pittsburgh, 100 Allen Hall, 3941 O'Hara Street, Pittsburgh, PA 15260, USA}

\author[0000-0002-2491-8700]{S. Alwin Mao}
\affil{Department of Physics \& Astronomy and Pitt PACC, University of Pittsburgh, 100 Allen Hall, 3941 O'Hara Street, Pittsburgh, PA 15260, USA}

\author[0000-0002-7918-3086]{Matthew W. Abruzzo}
\affil{Department of Physics \& Astronomy and Pitt PACC, University of Pittsburgh, 100 Allen Hall, 3941 O'Hara Street, Pittsburgh, PA 15260, USA}

\begin{abstract}

In this work, we investigate the properties of a population of cool clouds in simulated galaxy outflows. Using data from the CGOLS isolated galaxy simulations, we generate catalogues of $\sim 10^5$ clouds. We describe the impact of two different supernova feedback models -- a centrally concentrated starburst and disk-wide distributed star formation -- on the resulting cloud population. In both cases we find that the mass distribution function $dN/dM \propto M^{-2}$, in good agreement with model predictions of turbulent fragmentation.  We explore how cloud properties change with distance from the galaxy and find no qualitative distinction between the two feedback modes, although significant quantitative differences exist in attributes such as the total number of clouds, their densities, etc. We further show that both internal cloud velocities and cloud--cloud relative velocities are described well by properties of turbulent motion, despite significant bulk radial velocities. Finally, we investigate the distribution of cloud sizes in the context of recent theoretical arguments about cloud survival in winds. We find that proposed cloud survival criteria are a good predictor of cloud survival, in both the case where clouds are primarily destroyed and the case where cloud growth occurs in the outflow.  

\end{abstract}

\keywords{Galaxy winds(626) --- Astrophysical fluid dynamics(101)}

\section{Introduction} \label{sec:intro}

Galactic-scale outflows are seen in many star forming galaxies throughout the Universe and at all redshifts \citep[e.g.][]{heckman1990,franx1997,pettini2001,shapley2003,weiner2009,steidel2010,george2014,venturi2024}.  In addition to their observed ubiquity, over the past several decades these outflows have come to be recognized as a critical component in theoretical models of galaxy evolution \cite[e.g.][and references therein]{somerville2015, naab2017, thompson2024}.  By removing gas from the interstellar medium (ISM) they can impact future star formation, explaining why, for example, the stellar masses of galaxies are lower than dark matter halo functions predict \citep[e.g.][]{larson1974, white1978, dekel1986, white1991}.  By transporting energy and metals into the circumgalactic medium (CGM), winds can alter the pressure and temperature balance of the CGM, in turn impacting the accretion of gas back into the ISM along with subsequent star formation \citep[e.g.][]{tremonti2004, hopkins2012, faucher-giguere2016, fielding2017}.  Thus, outflows are a critical component in the ``baryon cycle", without which star forming galaxies would become quiescent in a fraction of their star formation history.

A salient feature of these winds is their multiphase nature.  For example, the outflow in the nearby starburst galaxy M82 was first observed in $\rm{H \alpha}$ emission tracing warm gas moving at speeds of $\sim 1000\kms$  \citep{lynds1963}.   M82's wind has since been observed across the electromagnetic spectrum, from X-rays emitted by $> 10^6\Kel$ plasma \citep{watson1984, strickland2009, lopez2020}, to millimeter emission from $<100\Kel$ molecular gas \citep{walter2002, leroy2015, chisholm2016, krieger2021}, and every phase in between \citep{shopbell1998, lehnert1999, hoopes2005, westmoquette2009, veilleux2009, xu2023}.  A general picture that has emerged from these spatially resolved observations is that outflows appear to have a volume filling hot phase, carrying a majority of the energy, with embedded cool clouds carrying more of the mass.

A main driver of these outflows are supernova (SN) explosions that shock-heat and entrain gas from the ISM \citep{chevalier1985}.   Given that the resulting wind is expected to be hot, explaining the presence of comoving cool gas presents challenges \citep{zhang2017}.   One possibility is that regions of the hot phase cool to $\sim 10^4\Kel$ after becoming dense enough to trigger bulk thermal instability \citep[as a result of ``mass loading'', e.g][]{thompson2016, Schneider2018b}.  Another possibility is that the cool clouds are patches of ISM accelerated via ram pressure by the fast moving hot phase.  However, it has long been noted that the time scale for this acceleration is greater than that for cloud destruction due to hydrodynamic instabilities and shocks \citep[see discussion in][]{thompson2024}.   As a result, many ideas have been proposed to stave off cloud destruction during this acceleration, such as invoking magnetic fields to provide surface tension, but these only work under certain conditions \citep{maclow1994, fragile2005, shin2008, mccourt2015, banda-barragan2016}.

A more recent theoretical explanation for accelerating cool clouds, known as turbulent radiative mixing layer (TRML) entrainment, has focused on the rapid cooling that can happen at the boundary layer between phases \citep{begelman1990, gronke2018, fielding2020, fielding2022, Abruzzo2022, abruzzo2023a}.  If mixed, intermediate-temperature gas can cool faster than the rate at which hydrodynamical instabilities disrupt the cloud, and the cloud can gain advected mass and momentum from the hot phase \citep{armillotta2016, gritton2017, li2020, sparre2020, kanjilal2021}.  The condition for cloud survival can be recast into a minimum size requirement that clouds must meet to guarantee growth \citep{gronke2018, farber2022, abruzzo2023b}.

Most cloud survival studies are undertaken in idealized environments, which has allowed the development of robust theoretical criteria given a relatively simple context.  However, these idealized environments are not fully representative of the actual conditions under which cool gas experiences acceleration in outflows, such as turbulent backgrounds, diverging flows, and temporally varying winds \citep{martizzi2020, Schneider2020, tan2023}. Thus, modeling populations of clouds in a global galaxy context can put these idealized small scale simulations into a new perspective. By comparing predictions made from idealized cloud-wind simulations with cloud properties obtained from a global galaxy simulation, one can evaluate how well these theories hold up in a realistic setting.  One can also test how population properties depend on global properties, such as the star formation rate or the star formation surface density.  

The Cholla Galactic OutfLow Simulations (CGOLS) project is an ideal testbed for such explorations \citep{Schneider2018a}.  CGOLS is a set of $5\pc$ resolution simulations designed to study M82-like outflows in a $2000\kpc^3$ volume. In this work, we explore cloud populations from the CGOLS IV \citep{Schneider2020} and CGOLS V simulations \citep{Schneider2024}, which differ primarily in the arrangement of the stellar clusters that drive the outflows.
This paper is structured as follows.  In \autoref{sec:methods} we describe the CGOLS simulations in more detail and explain how we generate our cloud catalogues.  In \autoref{sec:results} we illustrate properties of the cloud populations, including mass functions, physical properties as a function of distance from the galaxy, and inter- and intra-cloud velocity structure.  In \autoref{sec:discussion} we discuss our results in the context of previous work, including an exploration of the applicability of various cloud-survival criteria developed in smaller-scale simulations.  We close with a summary of our conclusions in \autoref{sec:conclusion}.

\section{Methods} \label{sec:methods}
\subsection{The CGOLS model} \label{subsec:cgols}

The cloud data used in this paper derive from two simulations that are part of the CGOLS project.  As mentioned in the introduction, the aim of CGOLS is to model supernova-generated multiphase winds across galactic distances ($\approx 10$ kpc), with constant parsec-level resolution. Each of the simulations begins with a rotating $10^4$ K gas disk with M82-like properties (gas mass $M_\mathrm{gas} = 2.5\times10^9\Msun$, gas scale radius $R_\mathrm{gas} = 1.6\kpc$, stellar disk mass $M_{*} = 10^{10}\Msun$, circular velocity $v_\mathrm{circ} \approx 130\kms$) embedded in a hydrodynamically stable hot halo \citep[see][for further details of the initial conditions]{Schneider2018a}. The box size is 10x10x20 kpc, with the galaxy located at the center, and the tall dimension aligned with the minor axis. The full volume is simulated at $\approx 5\pc$ resolution. 

In order to drive an outflow, discrete ``clusters" are added within the disk, at a rate consistent with the desired star formation rate. These clusters are individual spherical regions into which mass and thermal energy are injected at a rate set by theoretical models for the evolution of stellar populations \citep[e.g. Starburst99][]{leitherer1999}. CGOLS IV models SN feedback injection into the ISM from $10^{7}\Msun$ stellar clusters located randomly within $1\kpc$ from the galaxy center. We refer to this scenario in the present work as the ``central feedback" model. CGOLS V simulates a similar starburst galaxy, but with an exponential radial cluster distribution function that traces the disk gas and therefore extends to the edge of the domain (the scale radius in this case is $1\kpc$). We refer to this scenario as the ``distributed feedback" model. The cluster masses in CGOLS V follow a mass distribution function $\propto M_\mathrm{cl}^{-1}$, with low and high cutoff masses of $10^4$ and $5\times10^6\Msun$, respectively.

For the simulation snapshots analyzed in this work, we have assumed a constant star formation rate of $20 \mathrm{\Msun} \yr^{-1}$. We use a list of stellar clusters each with its own mass and orbit. Each cluster turns on when the cumulative star formation of the simulation equals the cumulative star formation represented by that cluster by order in the list. Active clusters provide feedback, which is modeled in both simulations as mass and (thermal) energy injection within a $30 \pc$ radius. Feedback is time-dependent, based on a starburst 99 model peaking at a cluster age of 3 Myr, half of the energy feedback deposited within the first $15 \Myr$, and nearly all of it within the first $40 \Myr$ \citep[i.e. a single burst model][]{leitherer1999}. In total, $10^{49}$ ergs and $0.18 \Msun$ are injected per solar mass. We refer the reader to the original papers for further details of the simulations \citep{Schneider2020, Schneider2024}.

\subsection{Cloud Identification} \label{subsec:cloudId}

A main goal of this work is to describe the properties of cool clouds embedded within the hot outflows generated in the CGOLS simulations. To do this, we have constructed a set of cloud catalogues for each simulation using the full grid snapshots, which are saved every 1 Myr. We define a cloud as a contiguous set of cells satisfying the temperature criterion $T < 2 \times 10^{4} K$. Cells are considered contiguous if they are adjacent or directly diagonal, so that a cell can be connected with any of its 26 neighbors in a $3\times3\times 3$ neighborhood.

To identify connected cells, we adapt the connected component algorithm provided by the Scipy Python library  (scipy.ndimage.label). In cases where it is convenient or necessary to separate the simulation domain into subdomains, we run the connected component algorithm within a subdomain and then use a classical union-find algorithm to reconnect clouds severed by subdomain boundaries. For each cloud, we store sums of the following values: volume, mass, mass weighted positions ($mx$, $my$, $mz$), momenta ($mv_{x}, mv_{y}, mv_{z})$, and kinetic energy ($mv_{x}^{2}, mv_{y}^{2}, mv_{z}^{2}$). These values add linearly when reconnecting clouds severed by subdomain boundaries. With our final cloud catalogs, we use these values to compute cloud properties such as the center-of-mass position, center-of-mass velocity, average density, and internal velocity dispersion.
%% TODO: Not sure what to call mv^2. For now calling it the kinetic energy.

\section{Results}\label{sec:results}

%% Cloud distributions, and as function of distance

For consistency with previous CGOLS analyses, we begin by focusing on snapshots taken 30 Myr after feedback is turned on.  In CGOLS IV this corresponds to the end of a period of maximum star formation with a $\dot{M}_{SFR}=20\Msun\yr^{-1}$. At this time, including the disk, the full simulation domain has 90,542 identified clouds with a combined mass of $1.9\times10^9\Msun$. Excluding the cloud associated with the disk, the combined mass is $1.6\times10^{7}\Msun$. At the equivalent time in CGOLS V, there are 385,285 clouds (including the disk) with a mass of $1.9\times10^9\Msun$. Excluding the disk, the combined mass is $1.4\times10^7\Msun$. In both simulations the cool clouds are embedded in a hot ($> 10^6\Kel$), tenuous, volume filling  phase, moving in galactic-scale biconical outflows.

A primary difference between CGOLS IV and V is the location of the cluster feedback; CGOLS IV only has feedback in the central kiloparsec, while CGOLS V has disk-wide feedback. As a result, there are far more clouds at large angles in CGOLS V \citep[see e.g.][Fig. 10]{Schneider2024}. To more cleanly compare the properties of clouds between the two simulations, we therefore narrow our focus in the following results to clouds within a $60^\circ$ opening angle biconical region.  Within this region, a sizeable fraction of the cool gas, extending several kpc from the galaxy plane, is connected directly to the disk. This creates an analysis challenge in that, while the cool ISM disk gas is obviously not part of an outflowing cloud, the connected components far from the disk are.  To separate outflow components from the disk, we regenerate the cloud catalogs only for the region within the bicone and $> 0.5\kpc$ from the disk midplane.  In this volume, CGOLS IV hosts 33,486 clouds with a combined mass of  $7.3\times10^6\Msun$, and CGOLS V has 118,691 clouds with a total mass of $2.8\times10^7\Msun$. Thus, within the biconical selection region, the outflow from the distributed feedback simulation has $3.8$ times more mass in $3.5$ times more clouds than the central feedback model.

\begin{figure*}
    \centering
    \includegraphics[width=0.48\linewidth]{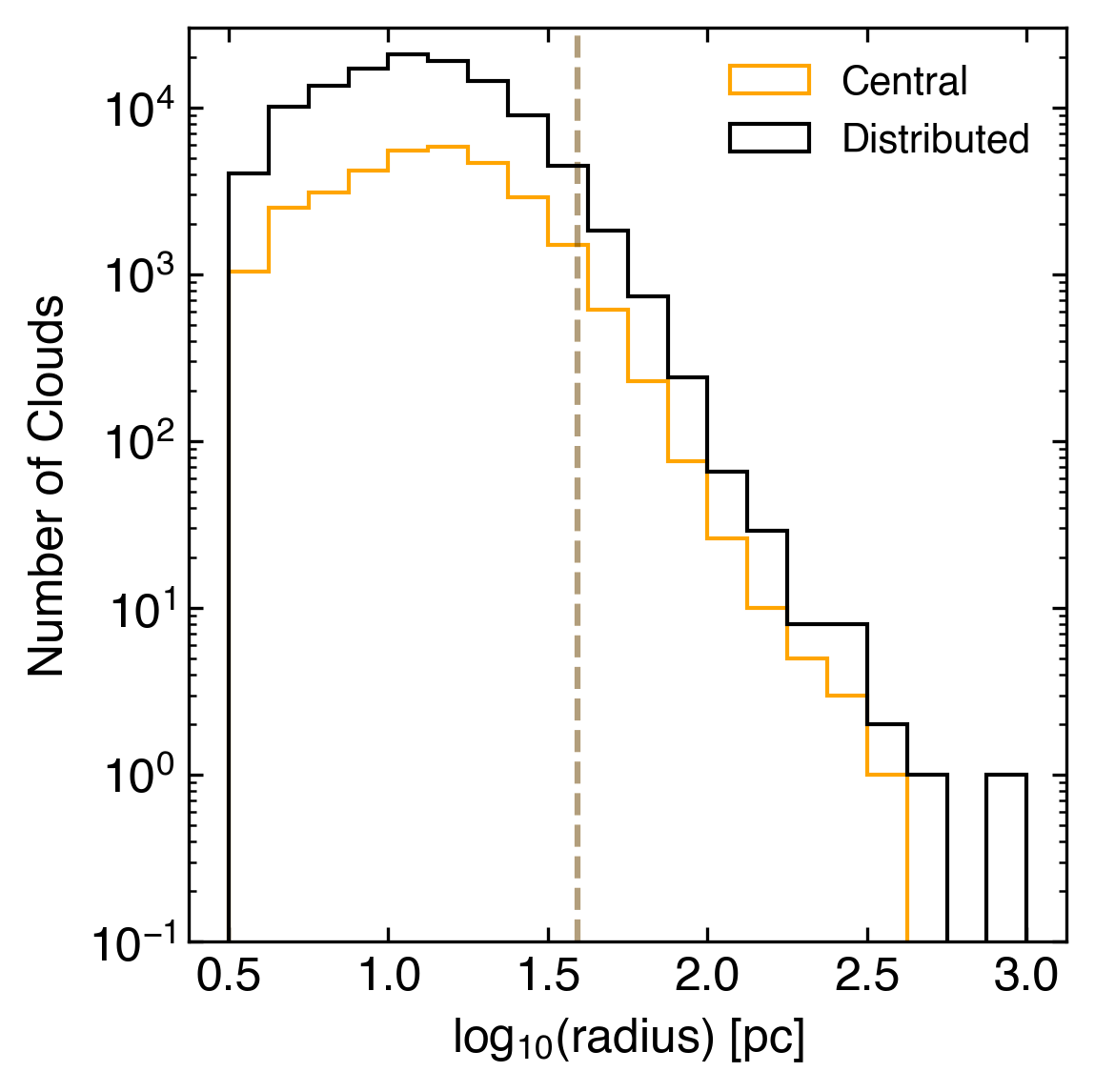}
    \includegraphics[width=0.48\linewidth]{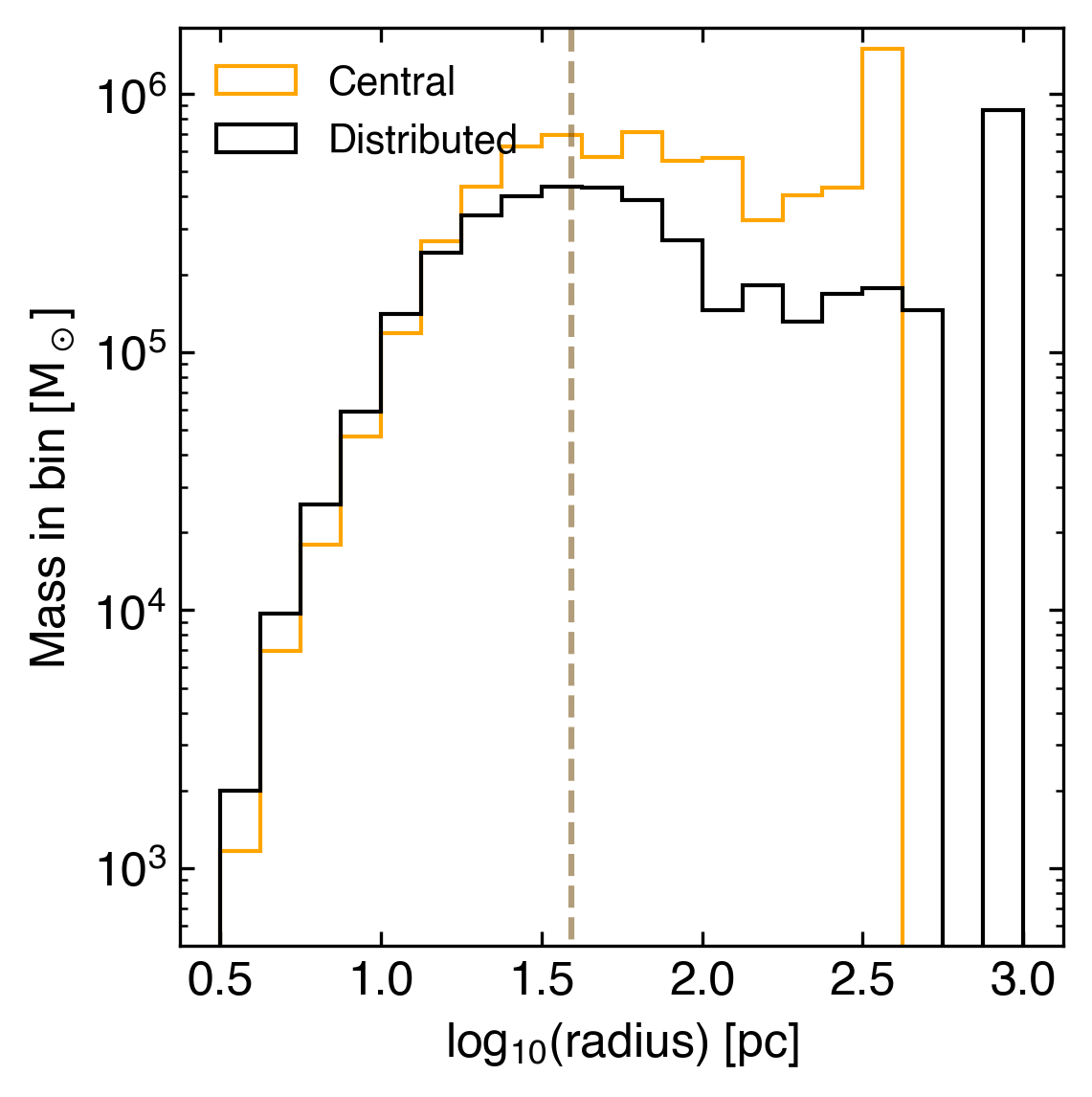}    \caption{Left panel: Number of clouds of a given radius. The grey dashed line at a radius of 40 pc marks the minimum size for a cloud to be considered ``resolved" ($8\Delta x$, with $\Delta x = 4.88$ pc). Right panel: The total mass contained in clouds of a given radius. Note that these plots do not include the two largest distributed feedback clouds (with radii $1.65\kpc$ and $1.46\kpc).$}
    \label{fig:r_histogram}
\end{figure*}

Despite this narrowed selection region, which explicitly excludes the ISM, two clouds in the CGOLS V catalog still have extreme properties. In the full simulation volume cloud catalog these two clouds are connected directly to the ISM. However, in the regenerated catalogs, their centers of mass are located at heights $z=-3.5\kpc$ and $z=3.9\kpc$ from the galaxy plane, and their radial velocities are $v_r = 356 \kms$ and $v_r = 458 \kms$, demonstrating that these clouds are genuinely part of the outflow.  These clouds account for $2.3\times 10^7 \Msun$ (over $80\%$ of the total cool mass in the catalogue). The CGOLS IV catalog also contains two outliers on either side of the disk that are associated with the ISM in the full volume catalog. However, in the CGOLS IV case these outliers are not as extreme. The largest cloud, at $z=2.0\kpc$, has a mass $1.5\times10^6\Msun$ ($\sim20\%$ of the total), while the next most massive cloud, at $z=-2.4 \kpc$, contains $3\times10^5\Msun$.

\subsection{Cloud Sizes}

In this subsection we explore the distribution of cloud radii in the two simulations. Radii are calculated assuming spherical shapes for measured cloud volumes. Since many of the clouds are elongated, these radii are only approximate indicators of linear size. \autoref{fig:r_histogram} plots histograms of the number of clouds as a function of cloud radius (left panel), and the mass in clouds per radial bin as a function of cloud radius (right panel). Throughout this section, data from the central feedback simulation (CGOLS IV) is shown in orange, while data from the distributed feedback simulation (CGOLS V) is shown in black. In both simulations, the cloud radius distribution by number peaks at $\sim 15\pc$ (roughly 100 cells), though we note that this size is smaller than what we consider a ``resolved" cloud ($40\pc$, see below). The mass-weighted peak radius is slightly larger, around $\sim 40\pc$ (excluding the several large outliers discussed in the previous section). The mass-weighted median radius for CGOLS IV is $\sim 80\pc$, while for CGOLS V it is $\sim 1500\pc$. Removing the few largest clouds we find a similar value for both simulations; the mass-weighted median radius in both cases shifts to $\sim 60\pc$. Thus, even in the case where we exclude the most massive clouds, the majority of the mass in both simulations is in clouds we consider resolved.

As the left panel illustrates, there are roughly $\sim 1/3$ as many central feedback clouds across all radius bins.  The right panel shows a more uneven distribution of binned mass.  CGOLS IV clouds with radii $<10\pc$ have slightly less binned mass than do CGOLS V clouds, and have more total mass in the larger clouds, up to radii $ \approx 350$~pc where the central feedback distribution ends (and excluding the very large outliers). Given that there are fewer total clouds in all radius bins in CGOLS IV, this indicates that on average, the central feedback simulation preferentially generates a more massive cloud population (for a given cloud radius). This then implies that clouds in the central feedback catalogue have higher densities than those in the distributed feedback simulation (as we will demonstrate explicitly in \autoref{subsec:radial_dep}). 

This difference can be understood in the context of the different outflow properties between the two simulations. In CGOLS IV, the cool phase pressure is higher at any given distance (than CGOLS V), so clouds at a given distance have higher densities. As we will show in \autoref{subsec:radial_dep}, the number of clouds in CGOLS IV also peaks at a lower galactocentric distance than in CGOLS V. Since pressure in the outflow also decreases with distance, this means that the median CGOLS~IV cloud is found in a higher pressure environment than the CGOLS~V clouds.

\begin{figure*}
    \centering
    \includegraphics[width=0.48\linewidth]{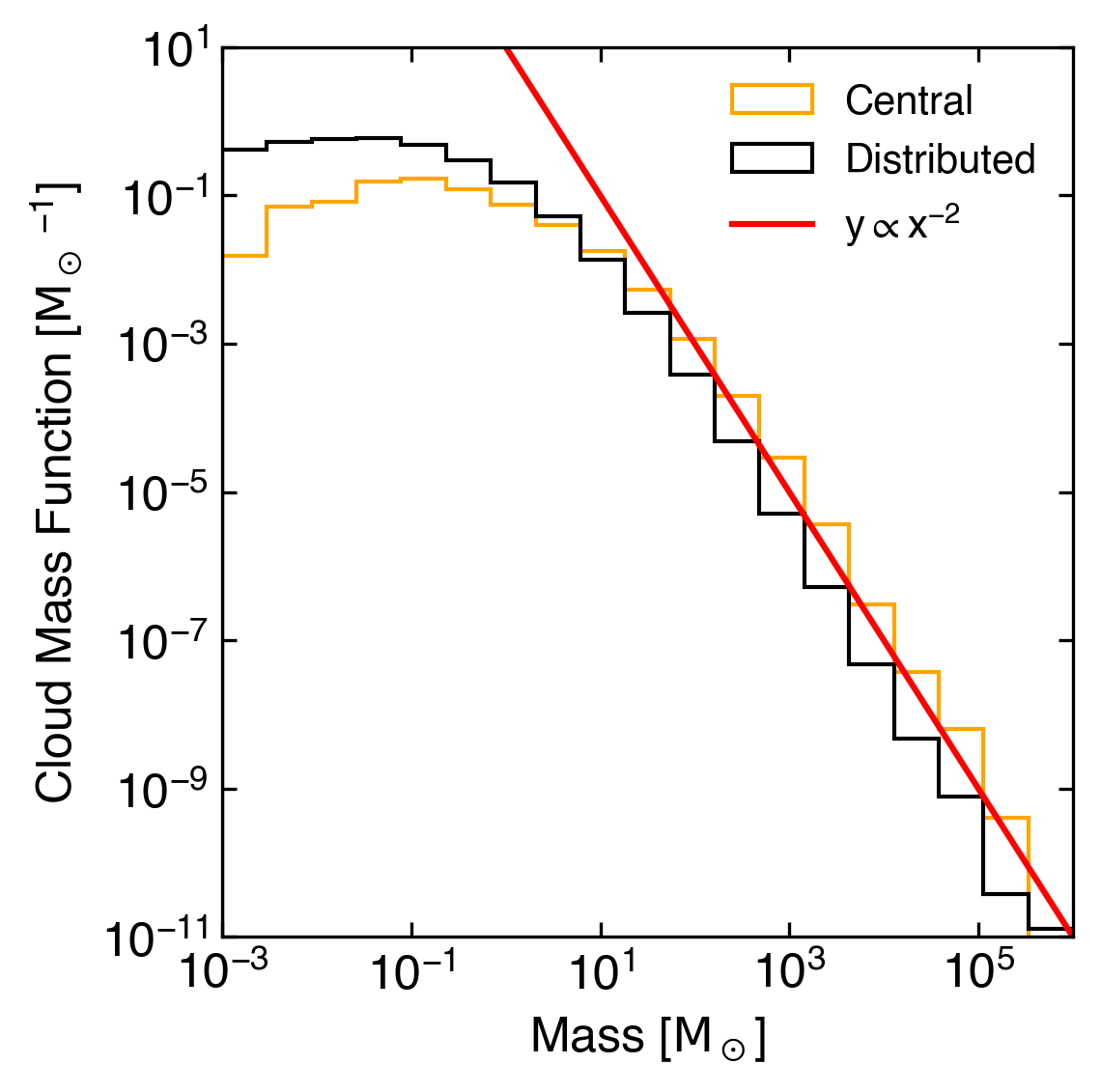}
    \includegraphics[width=0.48\linewidth]{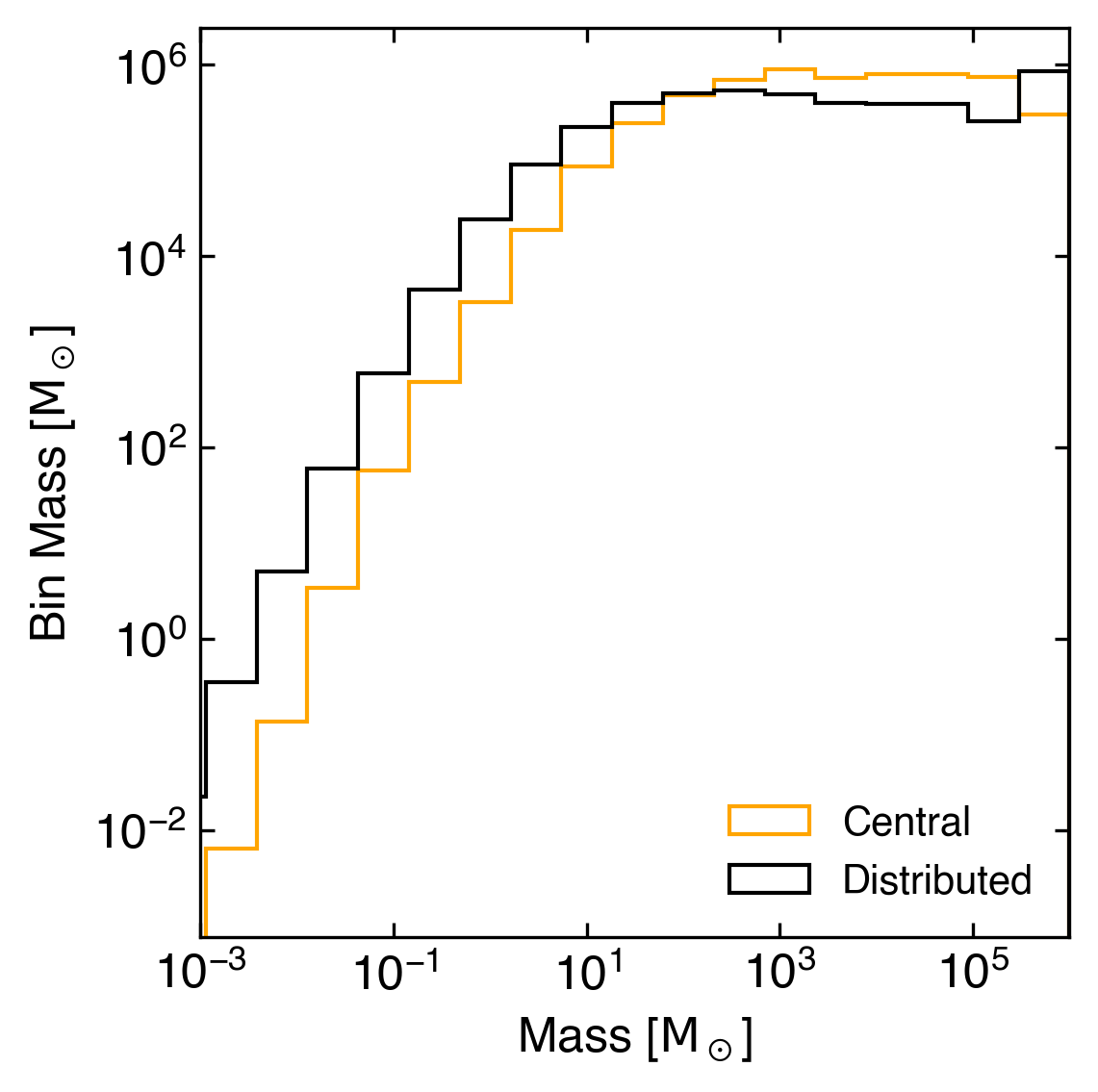}
    \caption{The cloud probability density distribution as a function of mass is shown in the left panel, along with a $-2$ power-law to guide the eye.  The mass-weighted histogram for the same data is on the right, showing a relatively flat mass distribution in both simulations above $\sim 10 \Msun$ and an excess of mass in the distributed feedback simulation at lower masses. Note that these plots do not include the two most massive distributed clouds, nor the single most massive central cloud.}
    \label{fig:mass_pdf}
\end{figure*}

In both panels we show a dashed line at $8\Delta x \approx 40\pc$, where $\Delta x$ is the resolution of the simulation grid. Clouds to the right of this line are larger than 8 cells in radius, and we deem them numerically resolved, based on results from idealized cloud-wind studies that demonstrate that certain aspects of cloud evolution (such as mass growth and destruction) converge at a resolution of 8 cells per cloud radius \citep[e.g.][]{gronke2020a}. This radial limit corresponds to $> 2140$ cells per cloud. We refer to clouds above this limit as ``resolved'' throughout this work. However, we note that because these radii are calculated using the cloud volume and assuming spherical symmetry, this definition of resolved does differ from that used in the idealized simulation setup.

\subsection{Cloud Mass Function} \label{subsec:mass_function}

\autoref{fig:mass_pdf} shows the distribution of cloud masses for both simulations, with a normalized histogram on the left, and a mass-weighted version on the right. In general, there are far more low mass clouds than high mass clouds, with the mass distribution in both simulations peaking at around $0.1\Msun$. In the left panel, we overplot a PDF, $dN/dM \propto M^{-\alpha}$, with $\alpha = 2$. This line is not a fit; we measure slopes of $\alpha = 2.00\, (2.01)$ for the central (distributed) feedback simulation for clouds with masses between $10- 10^5\Msun$, despite slight differences in their normalization. Thus, in both simulations, from $\sim 10\Msun$ to $\sim10^5\Msun$, the cloud mass distribution is described remarkably well by a power-law with a slope of $-2$. (The break off below this mass is likely due to insufficient numerical resolution, as we show in \autoref{sec:resolution}.) 

Such a relationship, over a wide range, is indicative of scale-free, turbulent fragmentation, a point we will return to in \autoref{subsec:velocity_struct}. Other recent works investigating the properties of clouds have found similar results. \citet{gronke2022} found the same power-law relationship for the mass distribution of droplets resulting from cloud fragmentation in turbulent boxes.  They explained the relationship as arising from the combination of proportional mass growth ($\dot{m} \propto m$) and droplet coagulation. Similarly, \citet{tan2023} found this relationship in the context of turbulence-driven tall box outflow simulations.  Since the relationship was seen at early times in their simulations, they reasoned that the scaling is associated with turbulent fragmentation of the ISM gas driven out during SN bubble breakout. \citet{fielding2023} also found this relationship in magnetohydrodynamic (MHD) simulations of thermally unstable turbulent systems. 

\begin{figure*}
    \centering
    \includegraphics[width=0.49\linewidth]{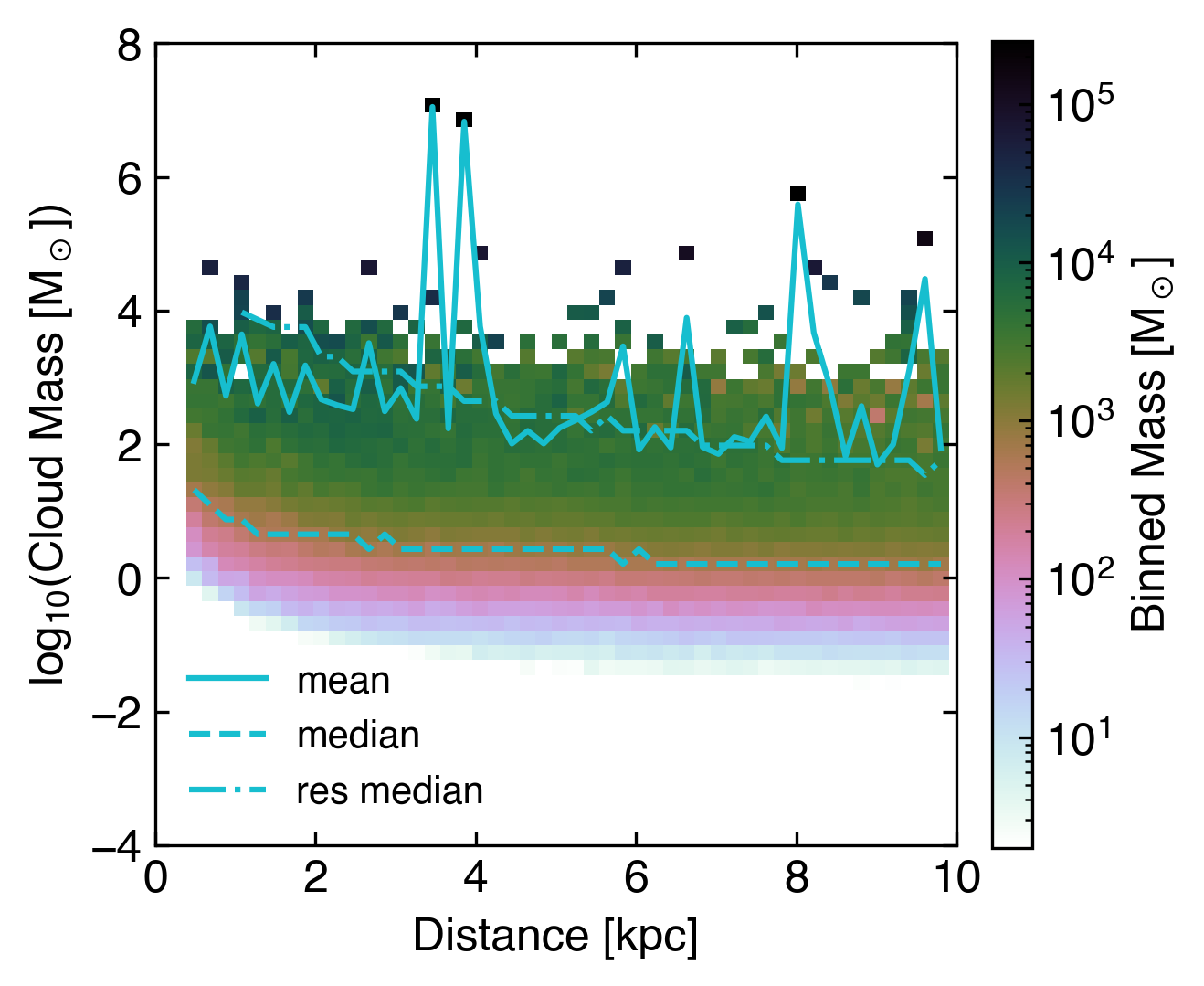}
    \includegraphics[width=0.49\linewidth]{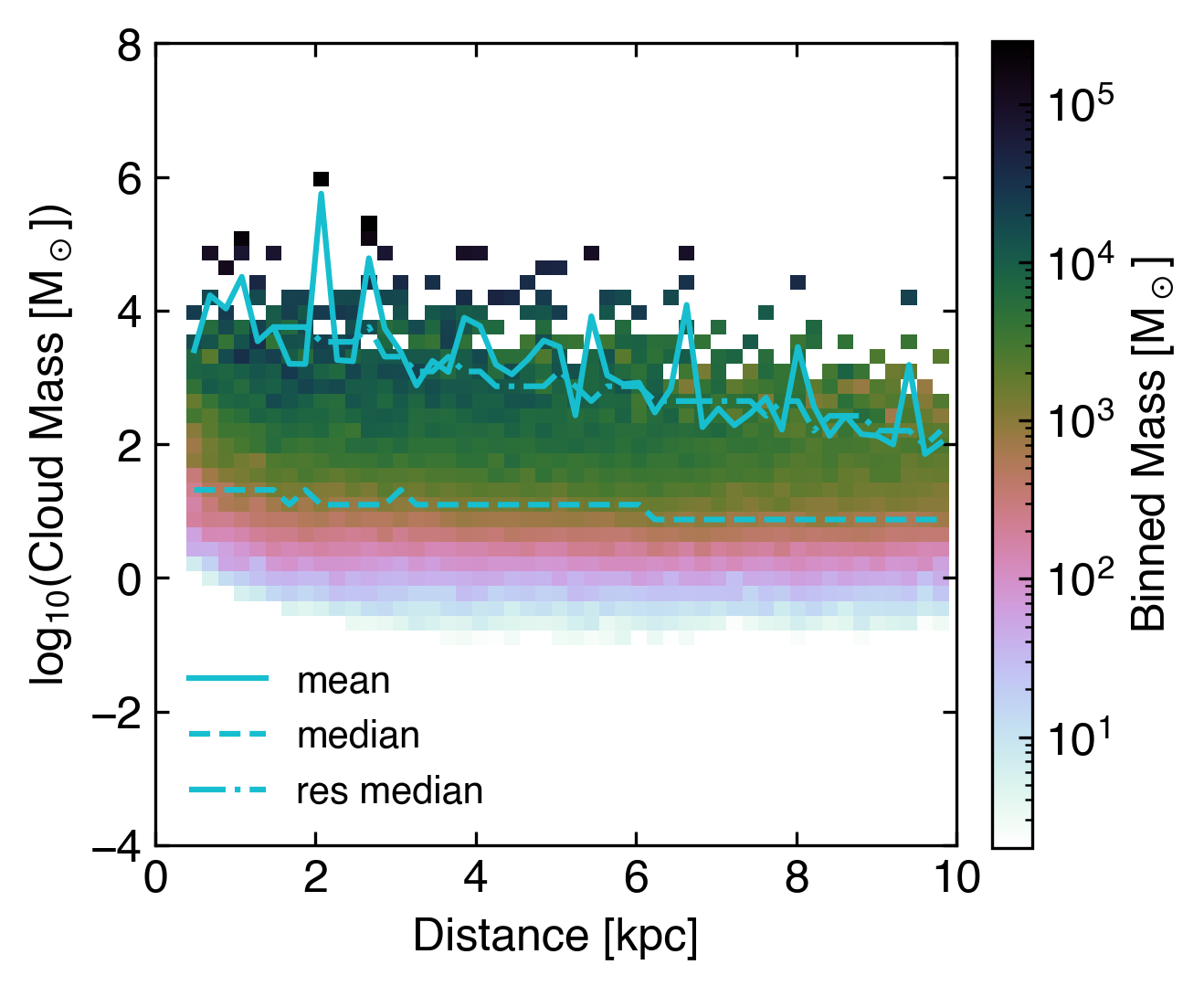}
    \caption{The distribution of cloud mass with respect to distance from the galaxy center, with  distributed feedback data plotted on the left, and central feedback data on the right.  The mean is heavily influenced by large mass outliers in both cases, but the overall trend revealed by the median in both simulations is towards decreasing cloud masses at larger distances.}
    \label{fig:2d_M_r}
\end{figure*}

A power-law slope of $-2$ corresponds to equal mass in equal logarithmic bins. In the right-hand panel, we see that this is not perfectly true, but over the upper range of cloud masses (above $10\Msun$), there is approximately equal mass in equal bins. This panel also elucidates the discrepancy between the similar total masses between the simulations, but the much larger number of clouds in the distributed simulation, since on the whole, the CGOLS V simulation has more lower mass clouds, and fewer high mass clouds, with the turning point at a mass of $\sim 10^2\Msun$. This may be partly a manifestation of the cloud locations (as described above), but it may also indicate that the outflow conditions in the distributed feedback simulation are more conducive to lower mass cloud production, which we discuss further in \autoref{subsec:R_crit}. However, we note that the two most massive clouds in the distributed simulation and the single most massive cloud in the central simulation are not included in \autoref{fig:mass_pdf}, and do not follow the same trends as the result of the cloud population.

\subsection{Distance Dependence} \label{subsec:radial_dep}

In this subsection we explore 2D histograms of various cloud attributes versus radial distance from the galaxy center. For each histogram we plot the mass weighted mean of the total cloud distribution at a given $r$, the unweighted median, and the unweighted median resulting from only selecting resolved clouds (where the cloud volume-based radii are $\geq 8\Delta x$). The mass weighted mean is calculated as the average, per radial bin, of the quantity being plotted multiplied by the mass in that bin, i.e. for the cloud radial velocities: $\langle v_r \rangle = \Sigma_\mathrm{N} (m_\mathrm{bin} v_\mathrm{r, bin}) / \Sigma_\mathrm{N} m_{\rm{bin}}$, where $m_\mathrm{bin}$ is the mass of clouds in the bin, $v_\mathrm{r, bin}$ is the radial velocity of the bin, and the summations are over all bins at that radius.

\autoref{fig:2d_M_r} shows the 2D distribution of cloud mass with distance, with the distributed feedback shown on the left and central feedback on the right. In both simulations, typical cloud masses (both the total median and resolved median) decrease with distance, which is consistent with either the destruction or fragmentation of clouds as they move out (or both). The mean cloud mass in both simulations also decreases with distance, though it is punctuated by large jumps caused by individual large clouds. For example, the two large ISM-connected clouds in the distributed feedback simulation can be seen at their mass-weighted mean position around 4 kpc, as can the single large ISM-connected cloud in the central feedback simulation at 2 kpc. The resolved median in both simulations is quite similar across the full range, dropping from $\sim 10^{4}\Msun$ at small distances, to $\sim 10^{2}\Msun$ by a distance of 10 kpc. In contrast, the total (resolved and unresolved) medians for the two simulations differ: at a distance of $\sim 10 \kpc$, the total median cloud mass is $\sim 10\Msun$ for central feedback, and $\sim 1\Msun$ for distributed feedback, though it is quite similar at small distances. In both simulations, high mass outliers dominate the average; an expected outcome given the power-law nature of the mass function.

\begin{figure}
    \centering
    \includegraphics[width=\linewidth]{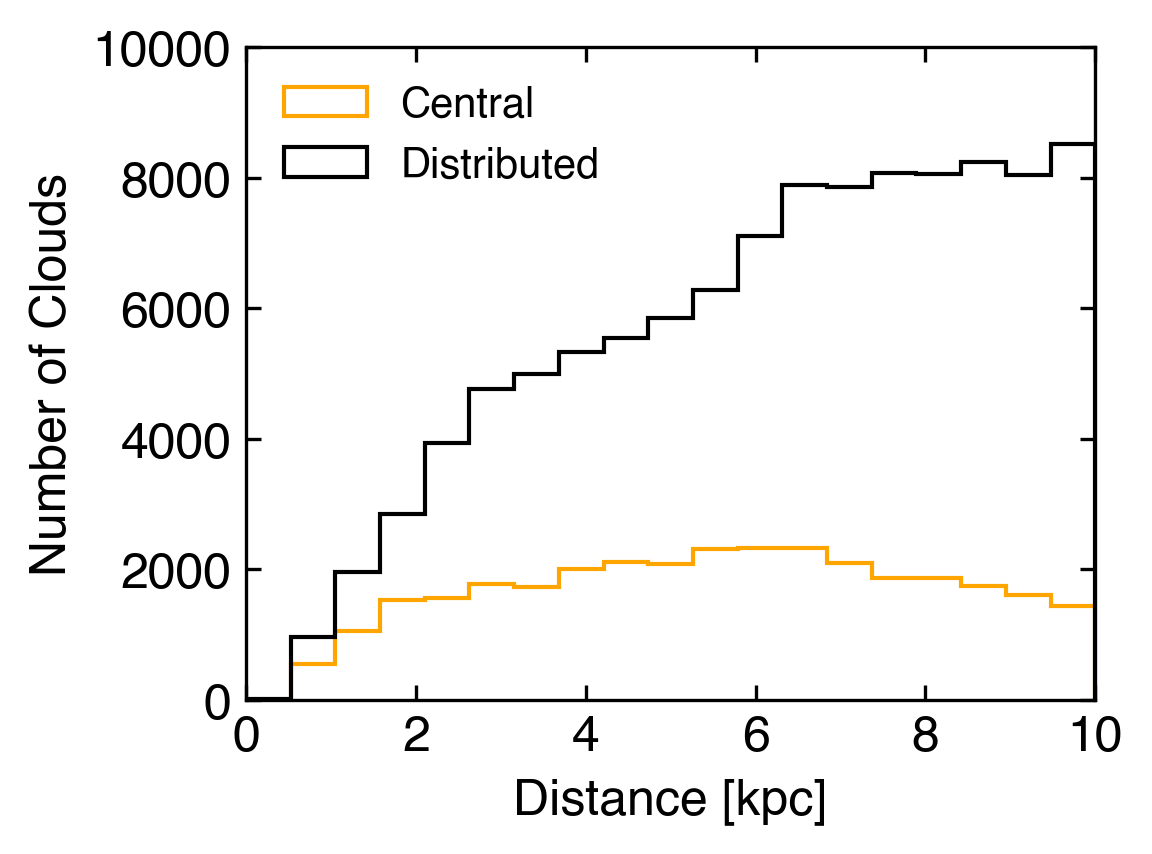}
    \includegraphics[width=\linewidth]{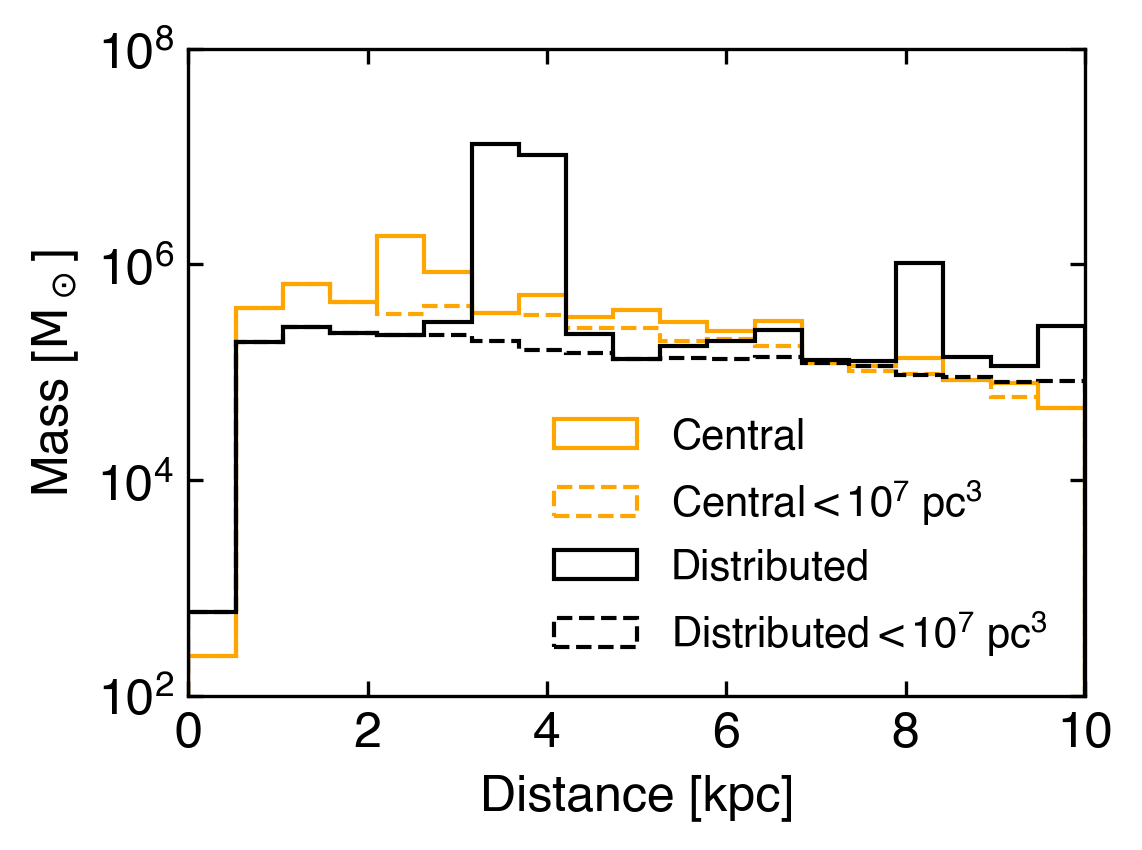}
    \includegraphics[width=\linewidth]{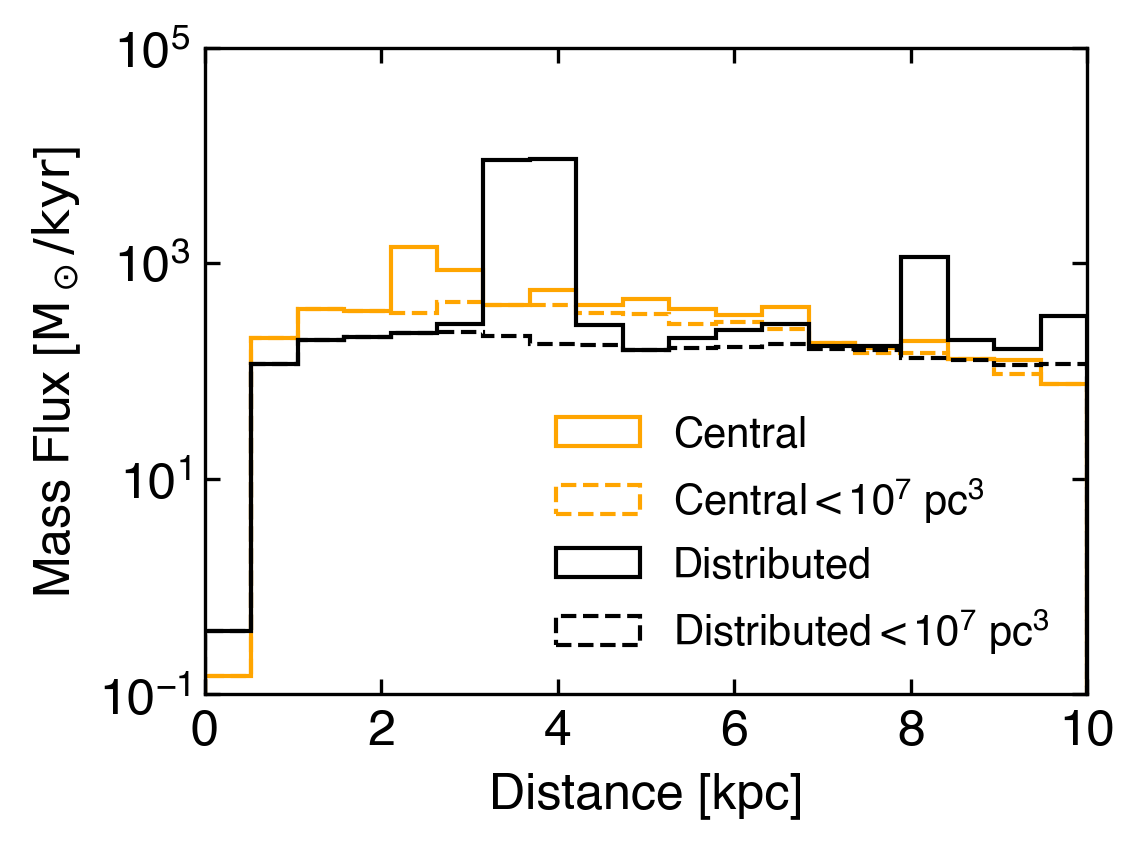}
    \caption{The histogram of the number of clouds versus distance from the galaxy (top), mass in clouds per distance bin (center), and the mass flux passing through the intersection of centered spherical shells and the outflow cone (bottom).  The number of clouds and total cool mass decrease with central feedback beyond $\sim 6 \kpc$.  The number of clouds increases with distance with distributed feedback, but the amount of mass per distance bin remains relatively stable. We also plot the resulting mass and mass flux when excluding clouds with volumes $>10^7\pc^3$.}
    \label{fig:numbers}
\end{figure}

To better understand why cloud masses (total per bin and individual) change with distance, in \autoref{fig:numbers} we show 1-D histograms of the number of clouds (top panel), mass in clouds (center panel), and cloud mass flux (bottom panel) as a function of distance. In the top panel, we see that in the central feedback simulation the number of clouds decreases beyond $\sim 6\kpc$, while in the distributed feedback case, the number of clouds continues to rise with distance. This is suggestive of net cloud destruction at large radii in CGOLS IV, but net cloud generation in CGOLS V. This does not necessarily translate to an increase in mass, however. In the central panel we see that the mass in clouds at any given radius decreases in both simulations (consistent with the averages plotted in \autoref{fig:2d_M_r}), albeit with large outliers in some bins due to individual large clouds in the distributed feedback case. To better visualize the trends for the majority of clouds, we also plot in dashed lines the mass per bin for a subset of the data that excludes the largest ($V > 10^7 \pc^{3}$) clouds, which removes 19 and 51 clouds from the central and distributed simulations, respectively. In this case, we see that in both simulations, the net trend is decreasing cool cloud mass with distance (per radial bin).

Total mass per distance bin may change with distance either due to net mass loss from the phase, or due to accelerating radial velocity. To disentangle these scenarios, we investigate the cloud mass flux (cloud mass times cloud radial velocity averaged over spherical shells), which is shown in the bottom panel of \autoref{fig:numbers}. To reduce the impact of high mass outliers, we again also plot the mass flux resulting from excluding all clouds with volume $>10^7\,\pc^3$. In the distributed feedback simulation, although \autoref{fig:2d_M_r} shows that the mass of individual clouds decreases with distance, \autoref{fig:numbers} shows that the net cloud mass flux is relatively stable with distance (punctuated with jumps from a few high mass clouds), which is suggestive of net cloud survival and fragmentation. By contrast, in the central feedback simulation, even when including the largest clouds, there is a slight decrease of mass flux with distance, indicating net mass loss from the cool phase.

\begin{figure*}
    \centering
    \includegraphics[width=0.49\linewidth]{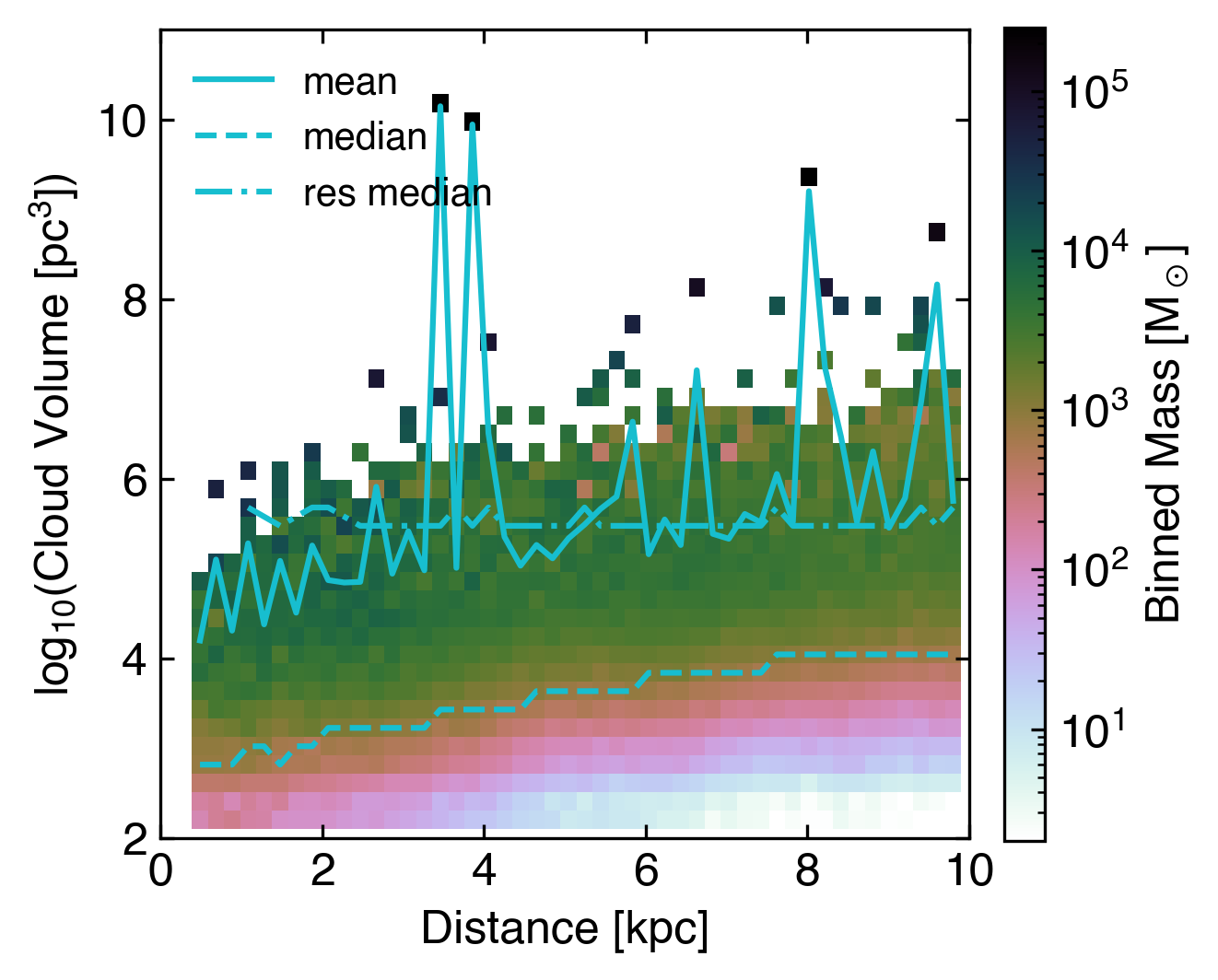}
    \includegraphics[width=0.49\linewidth]{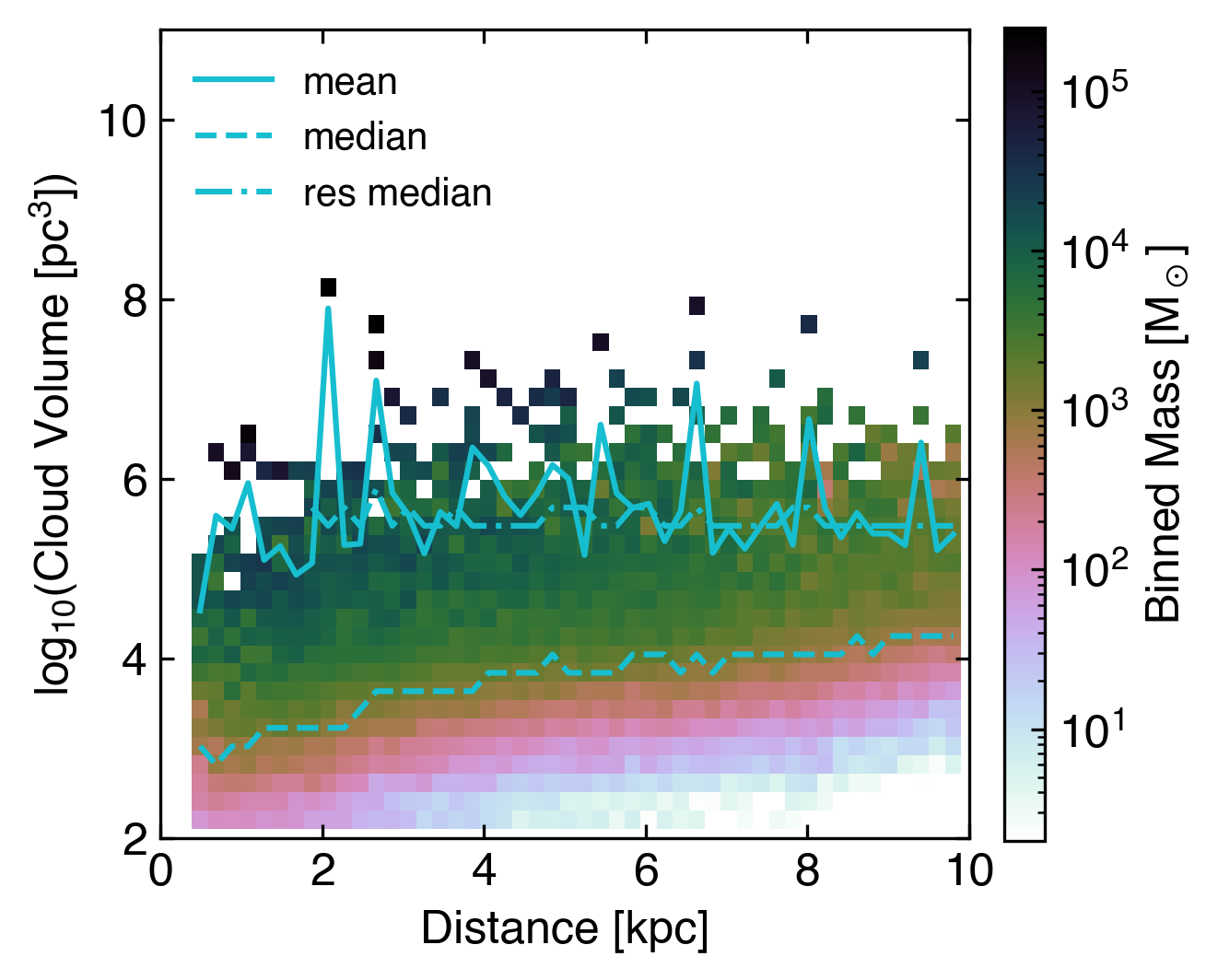}
    \caption{The distribution of cloud volumes with distance from the galactic center. Distributed feedback data is plotted on the left, central feedback data on the right.}
    \label{fig:2d_vol_r}
\end{figure*}

In \autoref{fig:2d_vol_r} we show 2D histograms of cloud volumes with distance.  In both simulations, the median resolved cloud volume is remarkably flat, at around $10^{5.5}\,\mathrm{pc}^3$. This may indicate a balance between two competing effects: fragmentation, which decreases the volume of individual clouds, and adiabatic expansion, which increases them. The total median cloud volume increases with distance, from around $10^3\,\mathrm{pc}^3$ to $10^4\,\mathrm{pc}^3$ over $10\kpc$, which likely reflects the decreasing pressure in the outflow in both simulations (leading to adiabatic expansion). The mean hovers around the resolved median, though it is again punctuated by jumps from large outliers -- with distributed feedback the two largest clouds have volumes $1.90\times10^{10}\pc^3$ and $1.30\times10^{10}\pc^3$, containing $1.29\times10^7\Msun$ and $1.03\times10^{7}\Msun$ of total mass, respectively. One other cloud has a volume $>1\kpc^3$, with mass $8.63\times10^5\Msun$. %, followed by four other clouds with volume $>10^8\pc^3$ (and combined mass $3.36\times10^5\Msun$), and 44 clouds with volume $>10^7\pc^3$ (and total mass $4.66\times10^5\Msun$). %Note that, since clouds are highly elongated, this does not imply an actual $0.5\kpc$ radius spherical cloud---volumes are exact, but radii are approximate. %
The central feedback simulation, on the other hand, has no clouds $>1\kpc^3$; the largest cloud in this simulation has a volume $1.72\times10^8\pc^3$ and mass $1.49\times10^6\Msun$.  %There are 18 other clouds with volumes $>10^7\pc^3$ and combined $1.16\times10^6\Msun$.  Not including the two largest distributed clouds, the 19 largest central clouds have a combined mass ($2.66\times10^6\Msun$) greater than the combined mass of the 49 biggest distributed clouds ($1.67\times10^6\Msun$). We also plot the volume resulting from dividing the median cloud mass by an analytical $\rho \propto 1/r^2$ fit to the cloud density.

\begin{figure*}
    \centering
    \includegraphics[width=0.49\linewidth]{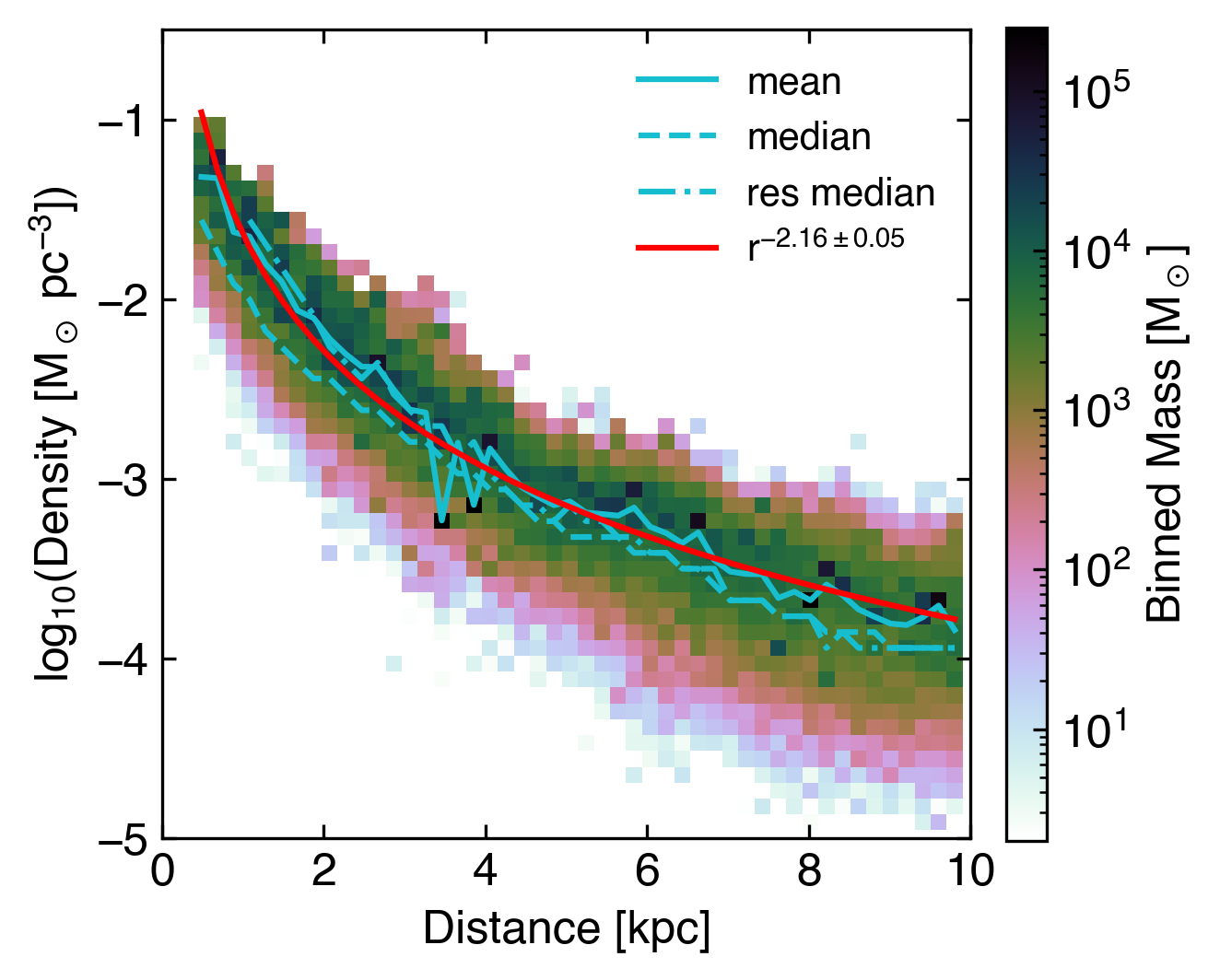}
    \includegraphics[width=0.49\linewidth]{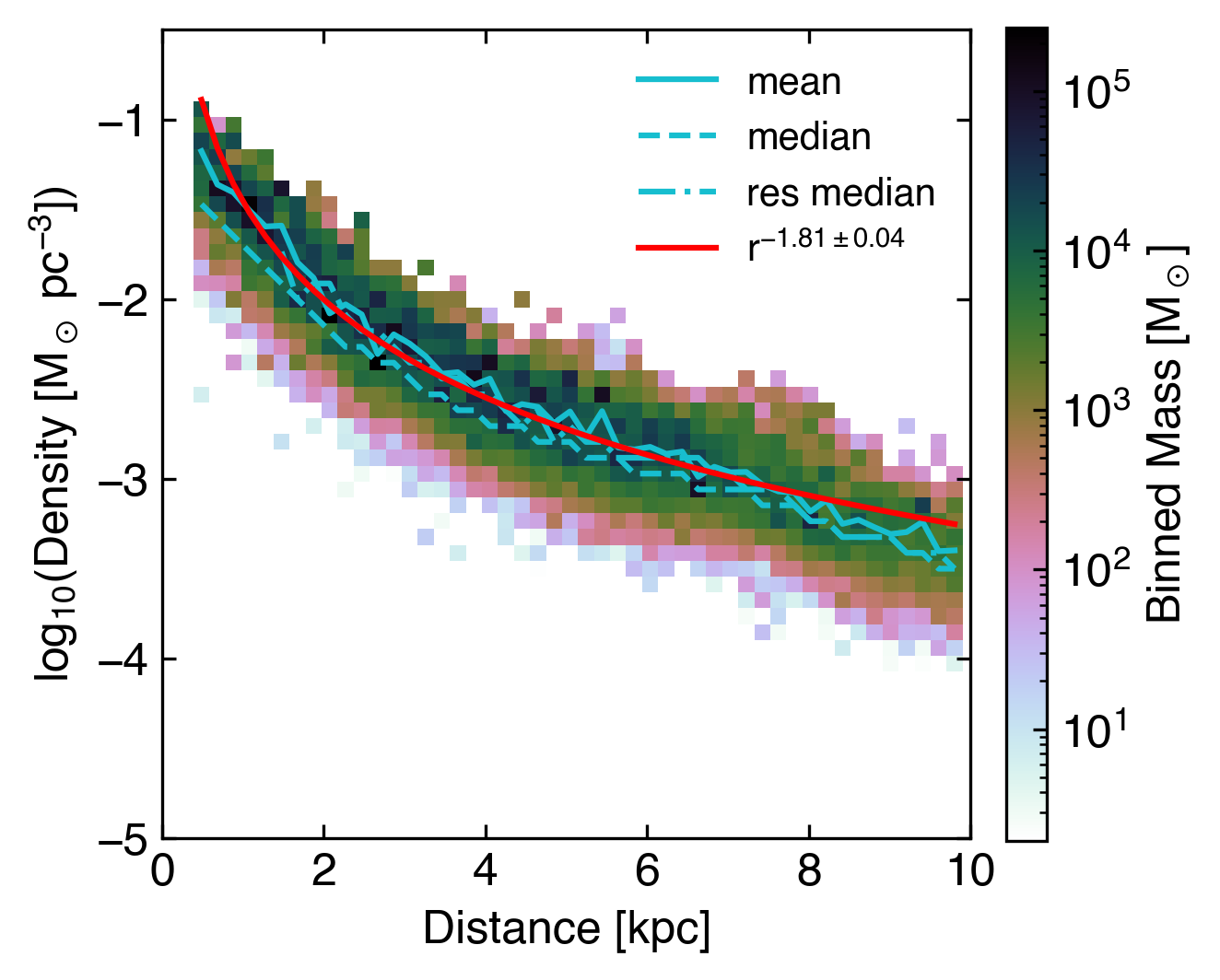}
    \caption{The distribution of cloud densities with distance from the galactic center.  Distributed feedback data is plotted on the left, central feedback data on the right.  We also show a $r^{-\alpha}$ analytic fit to the mean data.  Distributed feedback clouds tend to have lower densities at a given radial distance, and slightly steeper distance drop off. }
    \label{fig:2d_den_r}
\end{figure*}

In \autoref{fig:2d_den_r} we plot the cloud density distribution as a function of distance, along with power-law model fits to the mean shown in red. In both simulations, the cloud densities drop quickly with distance, as would be expected both due to adiabatic expansion, and as a reflection of the dropping pressure in the hot outflow confining them. Although the mean and median densities are quite similar between the two simulations at small radii, we see that the densities fall off more quickly in the distributed feedback case. This is reflected in the power-law fits -- the CGOLS IV clouds have a fit of $-1.81$, while for CGOLS V the fit is $-2.16$.

If the cloud velocities were constant and perfectly radial, we would expect an $r^{-2}$ dependence in steady-state, since in that case the mass flux through a solid angle $d\Omega$, $v_r \rho r^2 d\Omega$ must be constant. On the other hand, if clouds are accelerating as they move outward, we might expect a steeper profile. This is indeed seen in the fit for the distributed case but not with the central feedback model. This could be explained by net mass loss from the cool clouds into the hot wind, as suggested by the decreasing number of clouds and mass fluxes in \autoref{fig:numbers}. If higher density clouds at a given radius are preferentially surviving, the overall fit would have a flatter slope. We note, however, that the hot phase pressure is higher than the cool phase at all radii, so increased pressure due to interactions with the hot phase could also flatten the density slope.

\begin{figure*}
    \centering
    \includegraphics[width=0.49\linewidth]{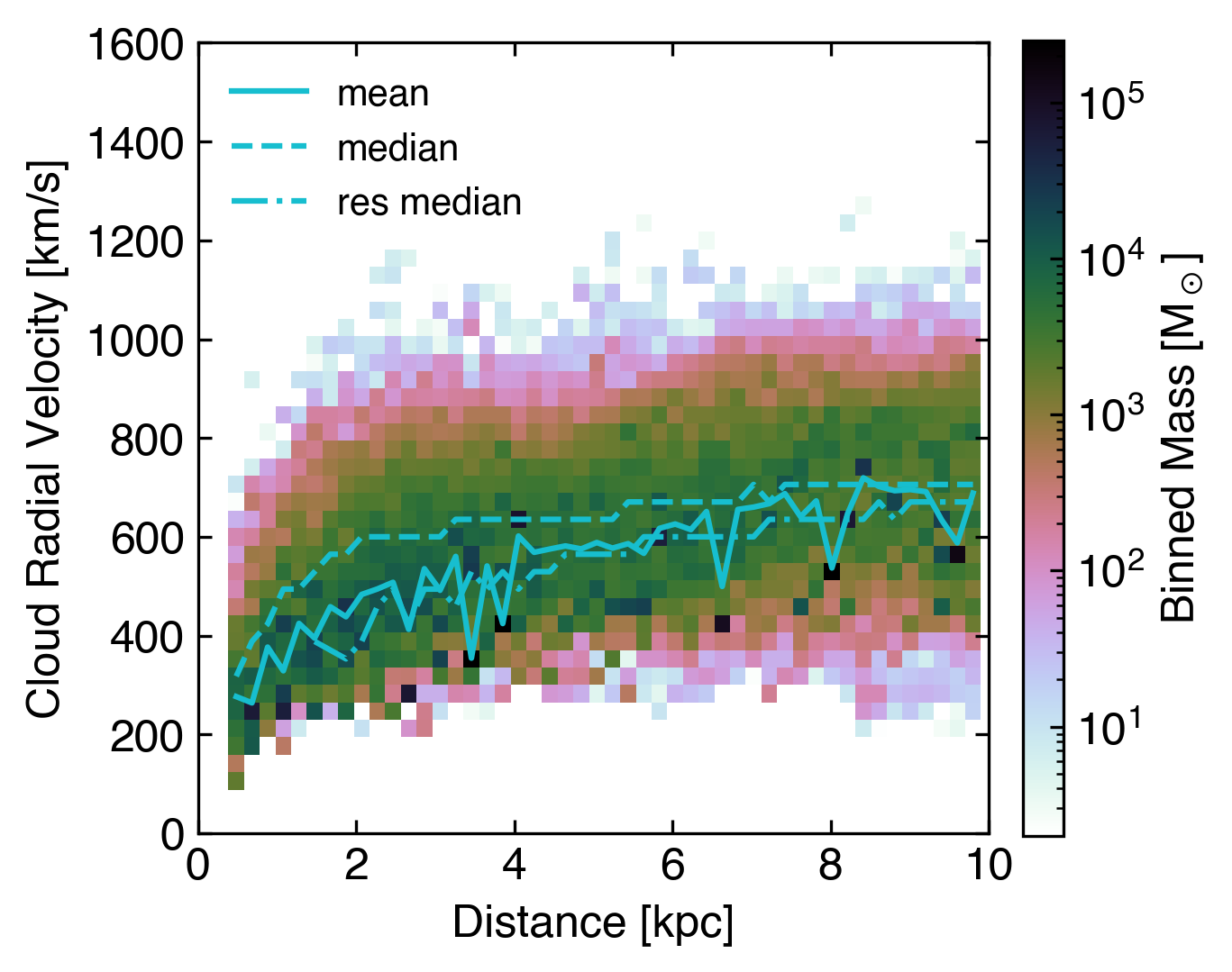}
    \includegraphics[width=0.49\linewidth]{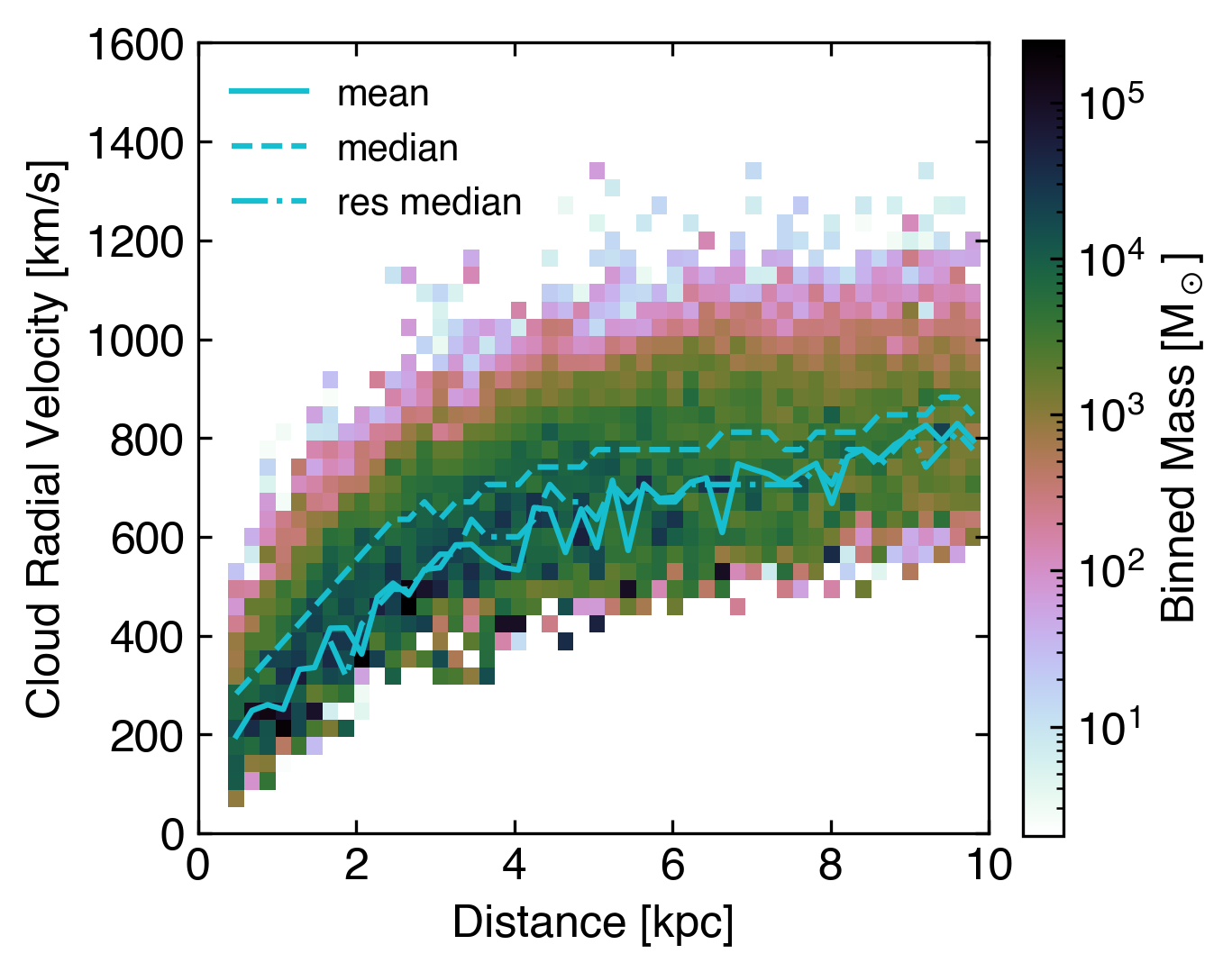}
    \caption{The distribution of cloud radial velocity with respect to distance from the galaxy center.  Distributed feedback data is plotted on the left, central feedback data on the right.  In both scenarios radial velocities rapidly increase with distance near the center, with the rate tapering off to a greater extent with distributed feedback.  Near $10\kpc$ clouds reach median speeds of $\sim 700\rm{km/s}$ with distributed feedback, and $\sim 850 \rm{km/s}$ with central feedback.}
    \label{fig:2d_vr_r_cone}
\end{figure*}

Figure \ref{fig:2d_vr_r_cone} shows how cloud radial velocity varies with distance.  In both simulations, the median radial velocity increases rapidly until $\sim 2-3\kpc$, beyond which the rate of increase slows significantly for both, though it remains slightly higher in the central model.  Resolved median speeds approach $\sim 775\kms$ at $10\kpc$ in CGOLS IV, and $\sim 670\kms$ in CGOLS V. In general, both median and mean velocities are lower at all distances in the distributed feedback simulation.

Distributed feedback also leads to clouds with greater radial velocity scatter.  Part of the explanation for this is geometric: clouds that are launched from larger cylindrical radii in the galaxy disk will tend to have substantial tangential velocity components when observed from the galaxy center, and larger vertical versus radial velocities relative to clouds launched closer to the center.  Clouds launched in the central feedback simulation, on the other hand, will have larger radial velocity components, for similar kinetic energies.  For example, a cloud launched from the disk at $5\kpc$ from the center, with a vertical velocity of $850\kms$ at $10\kpc$, could have a radial velocity of $850\times cos(30^\circ) = 736 \kms$ as seen from the center, which is close to the $\sim 700\kms$ distributed feedback median value at that distance.

Figure \ref{fig:2d_vr_r_cone} also shows that individual large clouds have preferentially lower velocities. These can be seen both as dips in the mass-weighted mean, and by the locations of single high mass bins, which trace individual massive clouds. For example, the two extremely massive clouds at $\sim 4$ kpc in the distributed simulation have velocities of $\sim 350 - 400 \kms$, while the resolved median velocity is closer to $500 \kms$. This is likely due in part to the fact that much of mass in these structures is located at smaller radii where less acceleration has taken place (those two clouds in particular are connected all the way back to the ISM). However, we see a similar effect for other clouds further out, where the median velocity has been much flatter for several kpc, so this does seem to reflect the fact that these large structures are moving more slowly than the median cloud. Similarly, the median velocities of resolved clouds are always lower than the median velocity of the total distribution. This could be a result of greater acceleration for smaller clouds, as would be expected if most of their momentum is being transferred via mixing from the hot wind. It may also reflect the increased effect of the galaxy potential on the more massive structures.

\subsection{Cloud Velocity Structure} \label{subsec:velocity_struct}

In this subsection we investigate two additional aspects of cloud velocity structure -- the intra-cloud velocity dispersions, as well as the inter-cloud velocity dispersion. In large part, this exploration is driven by the expectation that despite large bulk radial velocities, much of the cloud velocity structure derives from turbulence in the outflow that is constantly being generated both by the interaction of individual superbubbles from star clusters as they break out of the disk at small radii, as well as by the interaction of the hot background wind with the embedded cool clouds throughout the outflow. The extent to which the cloud properties align (or do not) with expectations set by theories of turbulence can thus shed light on the role that turbulence plays in determining those properties.

\begin{figure*}
    \centering
    \includegraphics[width=\linewidth]{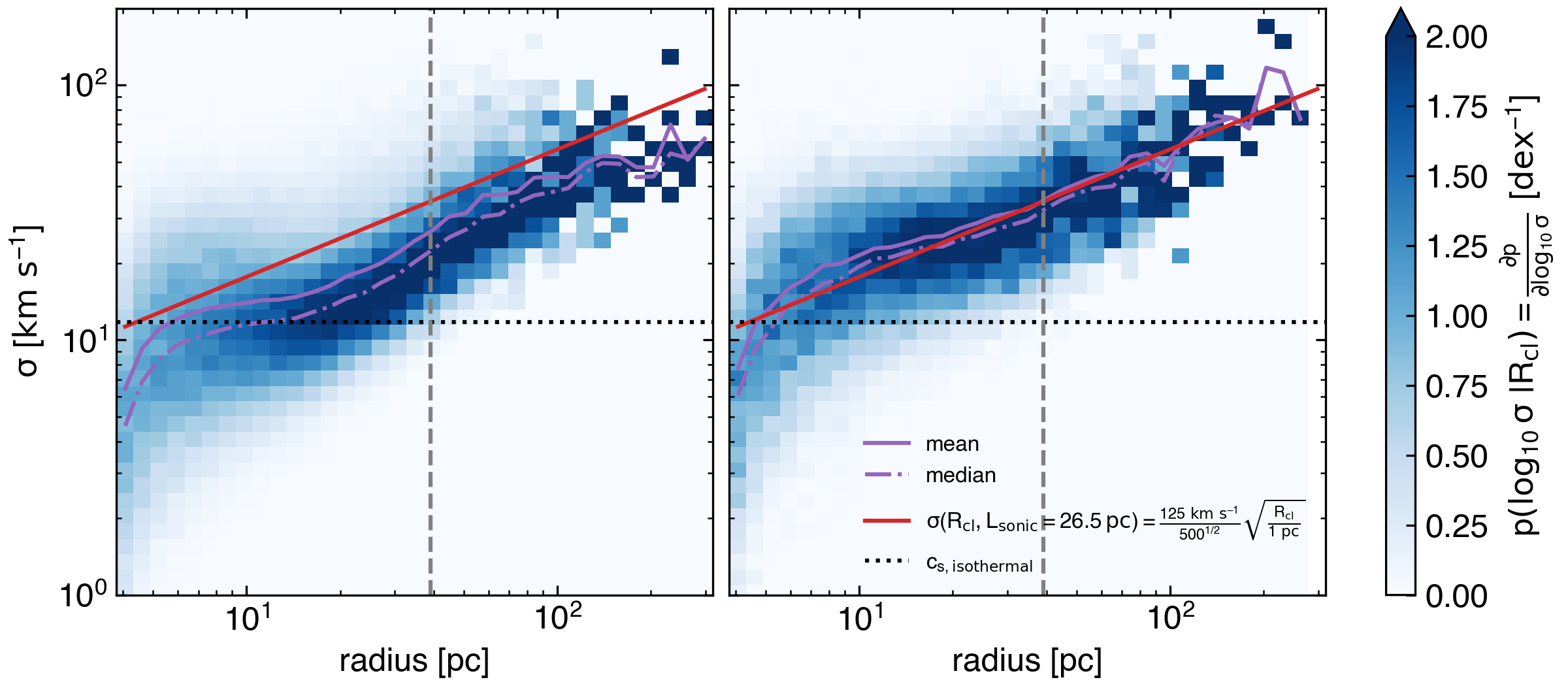}
    \caption{The conditional probability distribution function of the internal cloud speed dispersion for clouds of a given radius.
    Distributed feedback data is plotted on the left, central feedback data on the right.
    The vertical gray dashed line at a radius of 40 pc marks the minimum size for a cloud to be considered ``resolved'' ($8\Delta x$, with $\Delta x = 4.88\ {\rm pc}$), while the horizontal black dotted line specifies the isothermal sound speed.
    For reference, a version of Larson's Law is plotted in red: $\sigma \propto (\rm{radius/pc})^{0.5}\, \rm{km}\ \rm{s}^{-1}$. }
    \label{fig:2d_sigma_R_cone}
\end{figure*}

First, we explore the velocity dispersion of individual clouds. We define the internal cloud speed dispersion $\sigma$ as
\begin{equation}
 \sigma^2 = \sum_{i=1}^{3} \langle v_i^2\rangle - \langle v_i\rangle^2
\label{eq:sigma}
\end{equation}
where the averaging is over the cloud volume.  More specifically,
\begin{equation}
\sigma^2 = \sum_{i=1}^{3} \left(\frac{\int_V \rho(x) v_i^2(x)\, d^3x}{\int_V \rho(x)\, d^3x} - \left(\frac{\int_V \rho(x)v_i(x)\,d^3x}{\int_V \rho(x)\, d^3x}\right)^2\right).
\end{equation}
In other words, $\sigma$ is the velocity so that the internal kinetic energy is $\frac{1}{2}M\sigma^2$ for a cloud of mass $M$. This quantity is a measure of the cloud internal kinetic energy, which we plot as a function of cloud size in Figure \ref{fig:2d_sigma_R_cone}. This plot demonstrates that larger clouds typically have larger internal velocity dispersion and that the internal velocity dispersion of most clouds exceeds the isothermal sound speed. We now ask: does the latter property imply supersonic turbulence? 

To address this question, we consider Larson's law, an empirical relation for giant molecular clouds (GMCs) that describes a GMC's line of sight (LOS) velocity dispersion as a function of its LOS extent.
We adopt the fit from \citet{solomon1987}, $\sigma_{\rm LOS,GMC}(L)\approx (1\ {\rm km}\ {\rm s}^{-1}) \left(L/ 1\ {\rm pc}\right)^{1/2}$. When one assumes that $\sigma_{\rm LOS,GMC}$ doesn't include non-turbulent motions \citep[e.g.][]{draine2011}, the $L^{1/2}$ scaling can be considered evidence for supersonic, isothermal turbulence (i.e. Burgers turbulence).

Consider a turbulent engine that drives supersonic velocity differentials on length scales exceeding some quantity $L_{\rm sonic}$. The engine will not produce Burgers turbulence when $L_{\rm sonic}$ is large.\footnote{For example, velocity structure function measurements for gas from isolated clouds in Figure 17a from \citet{abruzzo2023a} have large-scale supersonic velocity differentials. These structure functions are more consistent with ordinary, subsonic Kolmogorov turbulence than supersonic Burgers turbulence.} Turbulent shocks are responsible for the properties of Burgers turbulence and these shocks require the turbulent engine to produce supersonic velocity differentials on a small spatial scale, $L_{\rm crit}$. Under the assumption that Larson's law describes Burgers turbulence, we infer a lower limit on $L_{\rm crit}$ by solving $\sigma_{\rm LOS,GMC}(L_{\rm sonic}) = c_{T,{\rm GMC}}$, where $c_T$ is the isothermal sound speed. We find $L_{\rm sonic,GMC} \approx 0.053\ {\rm pc}$ for $c_{T,{\rm GMC}} \approx 0.23\ {\rm km}\ {\rm s}^{-1}$\citep[the same value used to estimate $L_{\rm sonic, GMC}$ in][]{draine2011}.

We can construct a rescaled version of Larson's law, where $\sigma_{\rm LOS}$ equals isothermal sound-speed while holding at an arbitrary $L_{\rm sonic}$.
Accounting for $\sigma = \sqrt{3}\sigma_{LOS}$ (under isotropic turbulence) and $R_{\rm cl} \approx L/2$ yields
\begin{equation}
\label{eqn:Larson-rescake}
    \sigma(R_{\rm cl}, L_{\rm sonic}) \approx 125\, {\rm km}\ {\rm s}^{-1}
    \left(\frac{0.053\ {\rm pc}}{L_{\rm sonic}}\  
    \frac{R_{\rm cl}}{1\ {\rm pc}}\right)^{0.5}.
\end{equation}
Figure \ref{fig:2d_sigma_R_cone} illustrates a version of this relationship. The $R_{\rm cl}^{0.5}$ scaling is a remarkably good match to our clouds. However, it seems unlikely that our clouds undergo supersonic  Burgers turbulence given that they have $L_{\rm sonic}\ga 500 L_{\rm sonic,GMC}$.

\begin{figure*}
    \centering
    \includegraphics[width=\linewidth]{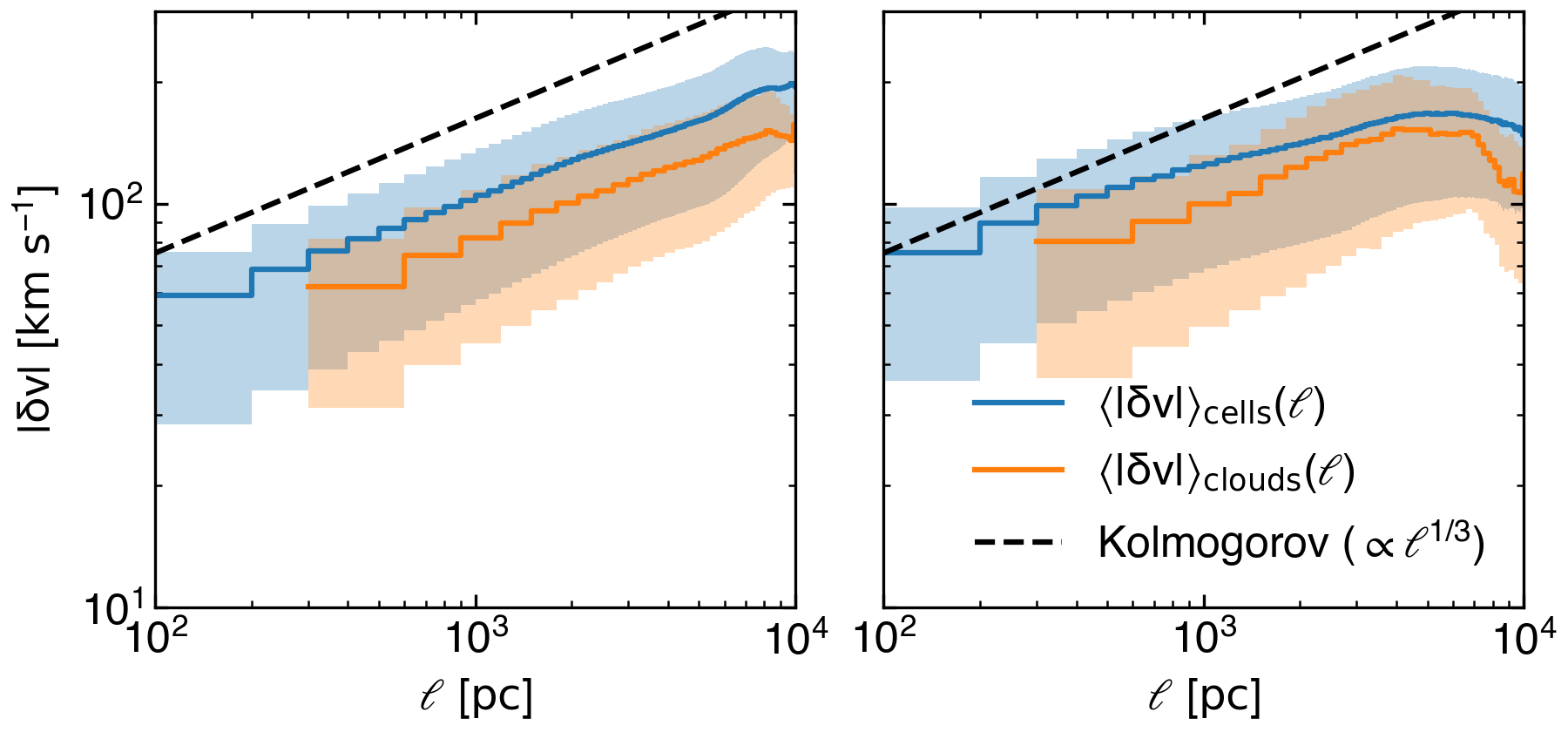}
    \caption{Solid lines show the first order velocity-structure function measured from bulk cloud properties, or the average velocity difference for all pairs of clouds in a given separation bin.
    The shaded region bounds the 25th to 75th quantile of the distribution of velocity differences.
    The left (right) panel shows data measured from distributed (central) feedback.
    The blue line estimates the structure function for all pairs of cells that compose the clouds; this calculation only includes clouds with radii smaller than 33.3 pc.
    The orange line shows the velocity structure function where each cloud is treated as an identical tracer of turbulence (i.e. velocity-differences are unweighted); these measurements include resolved clouds (radii of at least 40 pc) and omit clouds larger than 100 pc.
    The $\ell$ bin used in each calculation is 3 times larger than the maximum cloud size.
    The black dashed line illustrates the slope expected for subsonic Kolmogorov turbulence.} 
    \label{fig:vel_struct_bins2}
\end{figure*}

We highlight two caveats to these results.
First, our definition of $\sigma$ overestimates the dispersion from turbulence because it also includes contributions from coherent velocity gradients within the cloud \citep[e.g.][]{abruzzo2023a}. For example, we expect the front of a cloud to move at a different rate from the back, which gives rise to the clouds' characteristic head-tail morphology. The magnitude of this effect likely varies with how entrained a cloud is. Second, \autoref{fig:2d_sigma_R_cone} includes clouds from all distances, and it is plausible that the properties of turbulence may vary with distance in the outflow.
Nevertheless, the $R_{\rm cl}^{0.5}$ scaling remains a compelling result that deserves more consideration in future work.

%Another way of characterizing turbulent motion is through the power spectrum, which is the magnitude squared of the velocity in Fourier space. This is a measure of the power per unit volume and, like many quantities that vary over a wide scale, is assumed to take a power law form. %[Note: The large dynamic range is due to high Reynolds numbers ($\sim10^9$ in GMCs).  Although we do not include viscosity in our simulations, the grid length is our dissipation scale, resulting in a dynamic range $\sim 10^3$]. 
%For subsonic turbulence, Kolmogorov theory finds that the power spectrum, $P(k) \propto k^{-5/3}$, where $k$ is the wave number.
%The total power inside a region can be found by integrating the power spectrum from that physical scale --- say the size of the cloud --- to the smallest dissipation scale --- assumed zero if the dynamic range is large enough. 
%Since this quantity is also proportional to the speed dispersion squared, we can relate speed to the cloud size (i.e. Larson's Law), or inter-cloud distance: $\sigma_v \propto R^{1/3}$.
%In Burgers turbulence, supersonic motion creates shocks resulting in velocity step functions, whose Fourier transforms have a magnitude $\propto k^{-1}$, and a power spectrum $\propto k^{-2}$. In this environment, $\sigma_v \propto R^{1/2}$.

We now turn our attention to characterizing larger-scale turbulent motion, by measuring the first order velocity structure function for gas in the clouds. This is typically defined as the average velocity difference for all pairs of \emph{cells} separated by a distance $\ell$. We estimate this function, $\langle|\delta v|\rangle_{\rm cells} (\ell)$, by computing the weighted average of the bulk velocity differences between all pairs of \emph{clouds}. In order to account for the fact that we do not have cell-by-cell information, we weight each bulk velocity difference by the product of the corresponding two clouds' volumes. The products of two clouds' volumes is the product of the number of cells in each cloud (i.e. the total number of unique pairs of cells between two clouds, when each cell comes from a separate cloud) multiplied by the (constant) cell volume squared.
%We adopt this weighting scheme because the number of unique pairs of cells drawn from a pair of clouds (where each cell comes from a different cloud), is the product of the number of cells in each cloud and the number of cells in a given cloud is the cloud volume divided by the constant cell volume.

This is a good approximation when every pair of clouds satisfies two conditions. First, the bulk velocity difference of a cloud pair must approximate the average velocity difference between all pairs of cells from those two clouds.
Second, the separation in the mass-weighted positions of a cloud pair must be approximately the same as the separation between all pairs of cells in those clouds.
To ensure the latter condition is met, we ignore clouds with radii exceeding 33.3 pc and we use $\ell$ bins of 100 pc.

We also measure $\langle|\delta v|\rangle_{\rm clouds} (\ell)$, a version of the velocity structure function where we treat each cloud as an identical tracer of turbulence.
These calculations only consider resolved clouds (a radius of at least 40 pc) and don't involve any weighting.
The max considered cloud radius is 100 pc and the $\ell$ bins have a width of 300 pc.

\begin{figure*}
    \centering
    \includegraphics[width=0.48\linewidth]{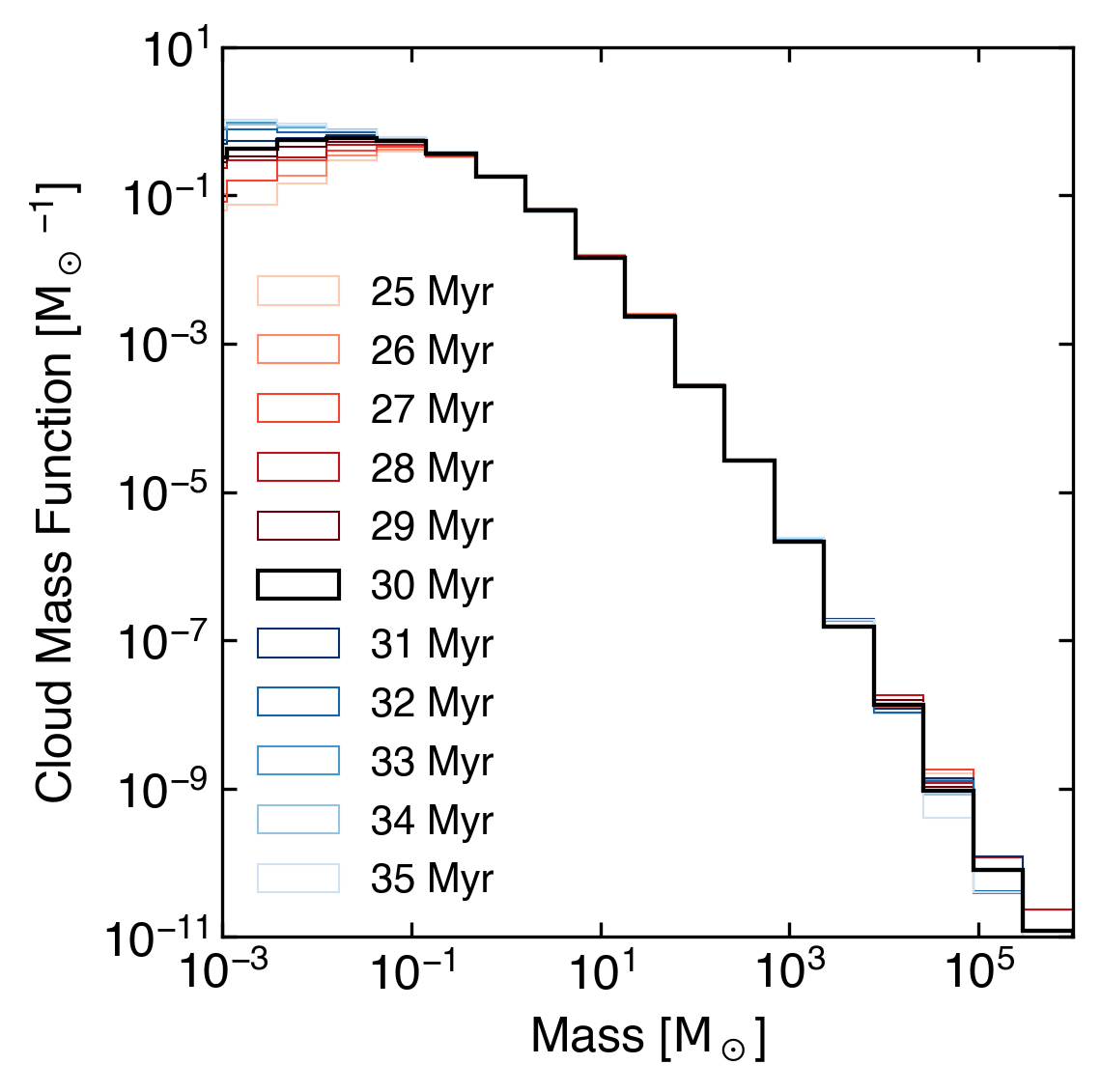}
    \includegraphics[width=0.48\linewidth]{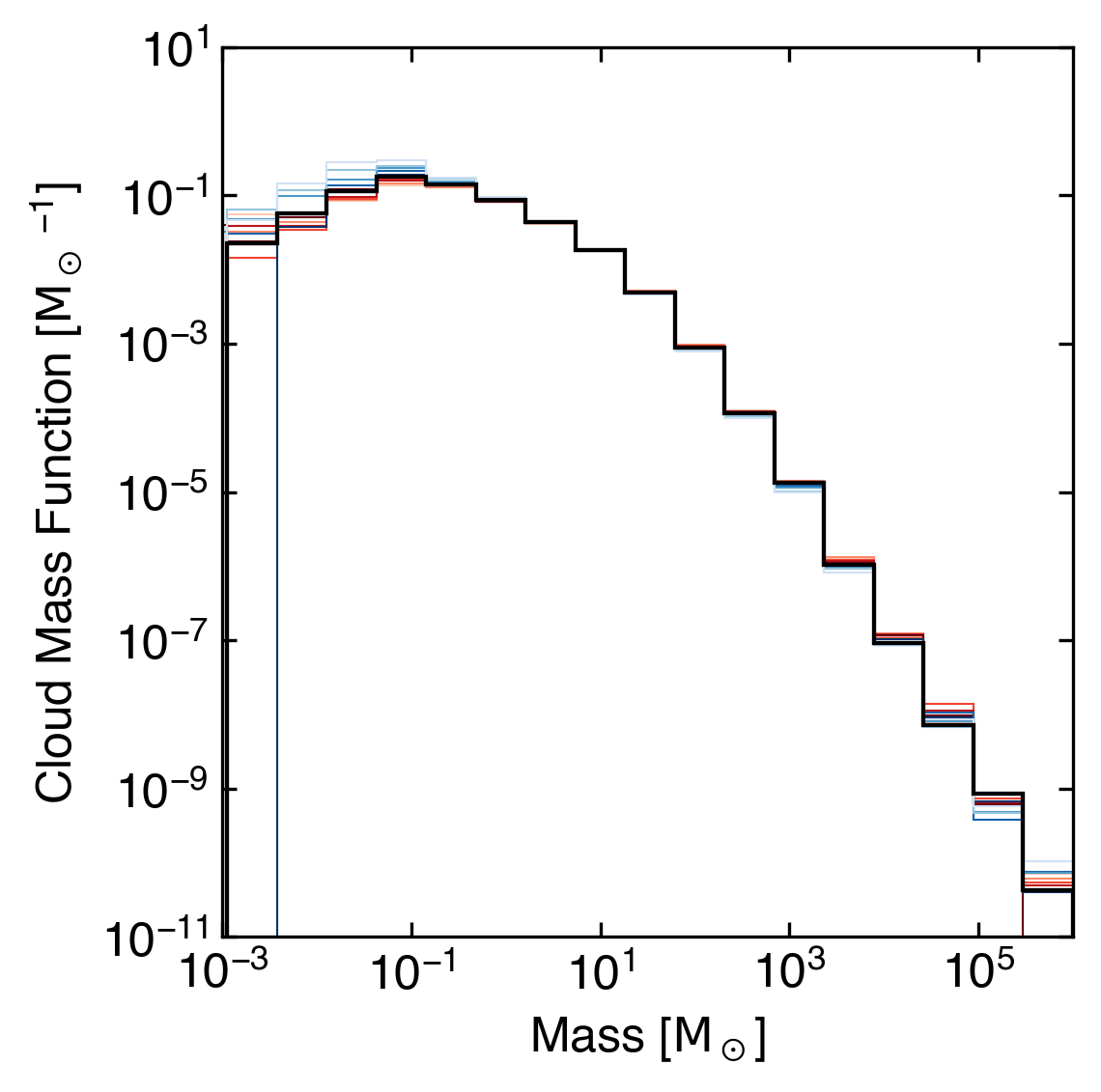}
    \caption{The cloud probability density distribution as a function of mass over the time interval $25-35\Myr$ for distributed (left panel) and central (right panel) feedback clouds.}
    \label{fig:time_mass_pdf}
\end{figure*}

In both cases we subtract estimates for the large-scale velocity from each cloud's velocity to reduce the impact of the large-scale velocity gradients. We get a smoothly varying estimate for the large-scale velocity by taking a moving weighted average of the radial component of the cloud velocity as a function of radial distance for clouds. The average uses a Gaussian filter with a standard deviation of 300 pc and we weight by cloud volume. Weighting by cloud mass gives qualitatively similar results.

\autoref{fig:vel_struct_bins2} shows $\langle|\delta v|\rangle_{\rm cells} (\ell)$ and $\langle|\delta v|\rangle_{\rm clouds} (\ell)$ for all clouds in the conical region above the midplane of the galaxy. Distributed and central feedback both produce $\langle|\delta v|\rangle_{\rm cells} (\ell)$ that act like 2-part broken power-laws. They respectively scale {\it roughly} as $\ell^{0.26}$ and $\ell^{0.30}$ for the smallest measured $\ell$ bins. While both scalings are somewhat shallower than the $\ell^{1/3}$ scaling expected for standard subsonic Kolmogorov turbulence, it's plausible that more robust estimates of the large-scale radial velocity may remove these differences.

Bigger differences manifest at larger $\ell$.
For distributed feedback, the slope slightly steepens at ${\approx}5\ {\rm kpc}$.
It is unclear whether this steeper slope at larger $\ell$ is an artifact of the domain-size or is a real feature.
In contrast, for central feedback, the structure function becomes constant at ${\approx} 1.5\ {\rm kpc}$.
This kind of change often occurs at the turbulent driving scale. This is noteworthy because the clusters in the central feedback model are confined to the central region of the disk with a diameter of 2 kpc. In other words, the structure function implies that in this simulation, the supernovae are responsible for driving large-scale turbulence.

\subsection{Time Evolution} \label{subsec:time_evolution}

Up to this point, we have only shown properties of the cloud population at a single snapshot in time, 30 Myr after the start of cluster feedback in each simulation. This analysis has been motivated by the fact that we expect the outflow driven by our constant star formation rate feedback model to be reasonably steady-state, once the initial shock has propagated out of the simulation domain. In this subsection, we test the validity of this assumption by investigating (a subset of) the previously explored cloud population properties as a function of time. To do this, we have compiled additional cloud catalogues for each simulation over the time interval $25-35\Myr$ using the method described in \autoref{subsec:cloudId}.  In both simulations the total amount of cold gas mass in the biconical selection region increases monotonically 3-fold over that time interval, but the overall trends remain the same.

\begin{figure*}
    \centering
    \includegraphics[width=0.8\linewidth]{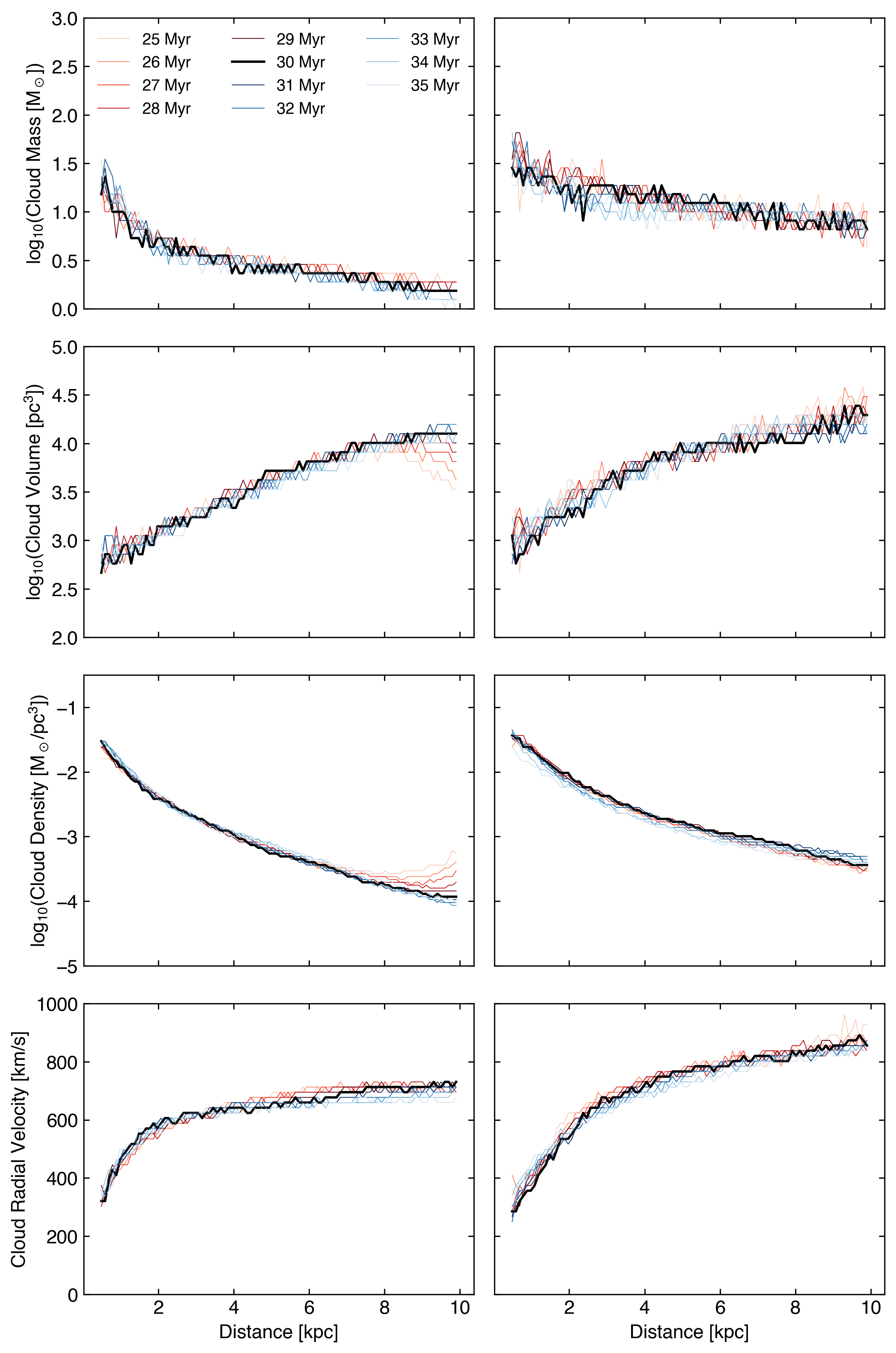}
    \caption{Median radial profiles for distributed (left) and central (right) feedback clouds over the time interval $25-35\Myr$.  Shown are the cloud mass, volume, density and radial velocity profiles plotted as functions of distance from the galaxy center (from top).}
    \label{fig:2dtime_median}
\end{figure*}

\begin{figure*}
    \centering
    \includegraphics[width=0.8\linewidth]{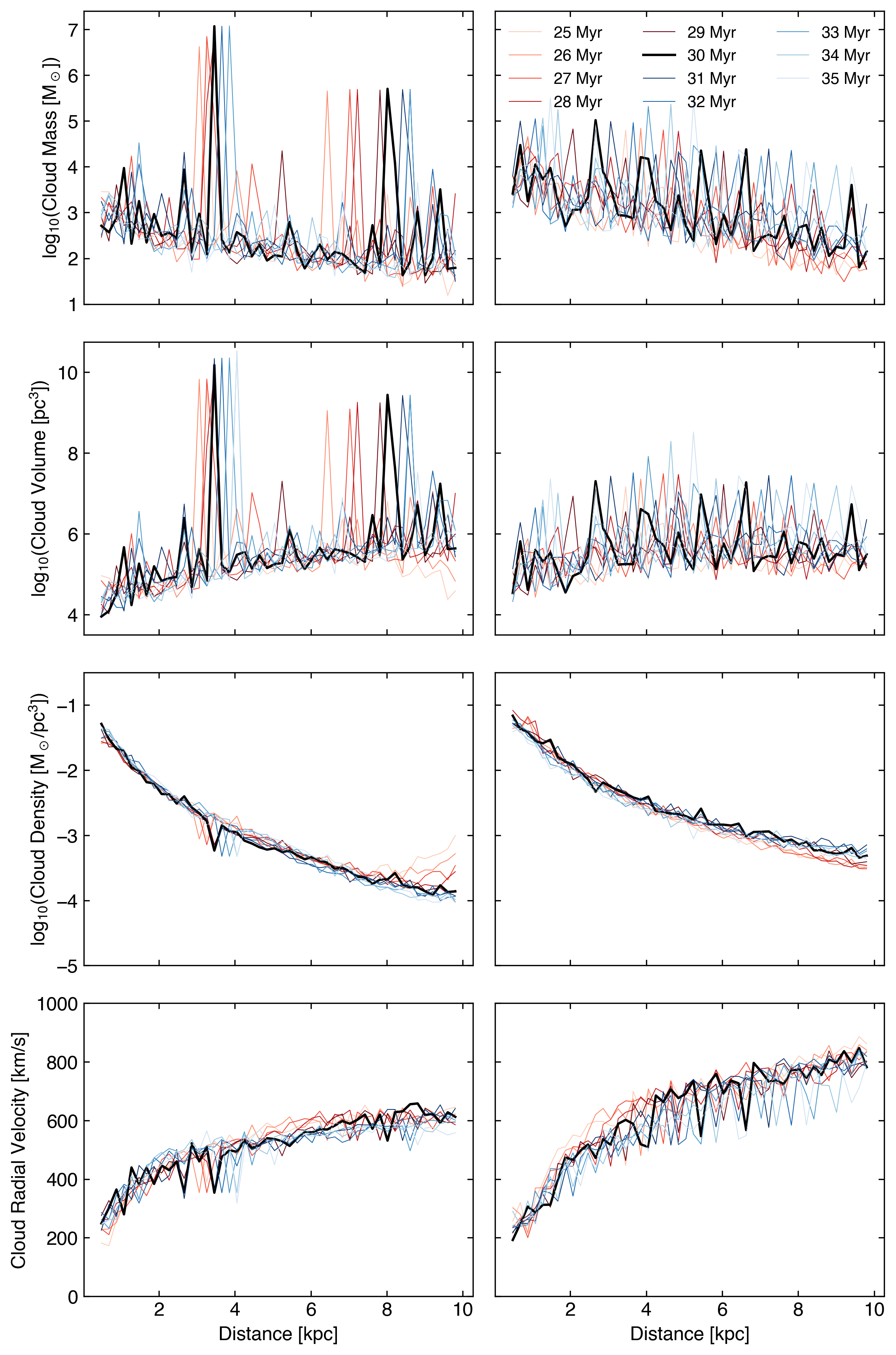}
    \caption{Like the previous figure, except that mean radial profiles are plotted over the time interval $25-35\Myr$.  For clarity, only clouds with negative $z$ values (below the galactic plane) are shown.}
    \label{fig:2dtime_mean}
\end{figure*}

\autoref{fig:time_mass_pdf} shows the cloud mass distribution function derived for each snapshot for the CGOLS V simulation (left) and the CGOLS IV simulation (right) over the time interval $25-35\Myr$.  The fiducial 30 Myr snapshot for each simulation is shown in black; these are the same lines that are plotted in the left panel of \autoref{fig:mass_pdf}. Red lines show times leading up to the 30 Myr point, while blue lines show the 5 Myr directly after. Although there are slight differences between the snapshots, particularly at the low and high mass ends of the mass function, the overall slope of the power law relationship is robust, as is the slight excess in low-mass clouds with distributed feedback relative to high-mass clouds with central feedback. Although we do not show them here, we have made similar comparisons for the other 1D histograms explored \autoref{sec:results}, and do not find significant differences. Thus, we conclude that deriving assumptions about the overall cloud population mainly from the 30 Myr snapshot is reasonable.

To further explore the time evolution of the cloud populations in the simulations, in Figure \ref{fig:2dtime_median} we show the evolution of the median radial profiles for cloud mass, volume, density and radial velocity during the same time interval, i.e. the median slopes from Figures \ref{fig:2d_M_r}, \ref{fig:2d_vol_r}, \ref{fig:2d_den_r}, and \ref{fig:2d_vr_r_cone}. Again, we see that the overall properties of the cloud population do not change substantially between snapshots. An exception is the median cloud volume and density in the distributed feedback simulation beyond $8\kpc$ at early times, before the simulation is able to reach steady state at large distances.  There also appears to be a slight shift in the median values of the cloud masses at small distances in the central simulation in the later snapshots. We note that this shift likely does represent a real change in the cloud population, as in this simulation the star formation rate drops from $20\Msun\,\yr^{-1}$ to $5\Msun\,\yr^{-1}$ during the 5 Myr after the 30 Myr snapshot, indicating that in this setup a higher star formation rate generates more massive clouds.

While the median profiles thus indicate that on the whole, the properties of the cloud population are stable as a function of time, an investigation of the mean profiles can provide an interesting view of how individual clouds (or collections of clouds) behave in the simulations over time. In Figure \ref{fig:2dtime_mean}, we show the average profiles for the same four quantities over the same time interval. In order to make it easier to track the evolution of individual clouds, we restrict the catalogue selection for this plot to only the bottom half of the simulation domain. Many of the same features can be observed in this plot as are seen in Figures \ref{fig:2d_M_r} - \ref{fig:2d_vr_r_cone}; in particular, the spikes caused by individual massive clouds are seen in each panel. Here, however, we are able to track the movement of individual clouds as they propagate through the domain. For example, the large, $10^{5.5}\Msun$ cloud that is located at $\sim 8\kpc$ in the 30 Myr distributed snapshot is seen to have moved a distance of $2\kpc$ over the course of 10 Myr. This distance is actually somewhat less than would be inferred from the radial velocity of the cloud, which is $500\kms$ -- at that rate, it should move $5\kpc$ over the given time interval. In addition, we can see that although this cloud appears to grow in volume, consistent with our expectation given the decreasing density, its mass remains roughly stable. This behavior perhaps indicates that while exceeding a specific size criterion is sufficient to predict cloud survival, other physics that is difficult to capture in wind tunnel simulations may play a role in determining cloud growth, such as fragmentation driven by interactions with a turbulent background. We explore this point further in \autoref{sec:discussion}.

\section{Discussion}\label{sec:discussion}
\subsection{Cloud Survival Criteria and Growth} \label{subsec:R_crit}

As discussed in the introduction, recent idealized cloud crushing and shear boundary simulations have shed light on how cool gas can be entrained in galactic outflows. These studies have yielded minimum cloud radius criteria required for cloud survival. These criteria, however, have not been evaluated in full galaxy simulations.

% Matthew's rewrite
We begin this section by elaborating on the cloud survival literature mentioned in the introduction. In what follows we consider a spherical cloud at rest, with radius $r_{\rm cl}$, density $\rho_{\rm cl}$, and temperature $T_{\rm cl}$. The cloud is embedded in a hot wind, with density $\rho_w$ and temperature $T_{\rm w}$ moving at a velocity $v_w$. All material is at some pressure $p_0$. Clouds avoid destruction and are successfully accelerated when the characteristic cooling timescale of the mixed gas is shorter than the relevant timescale describing the dynamics of the cloud-wind interaction. Under these conditions mass and momentum are transferred from the wind to the cloud.

\begin{figure}
    \includegraphics[width=\linewidth]{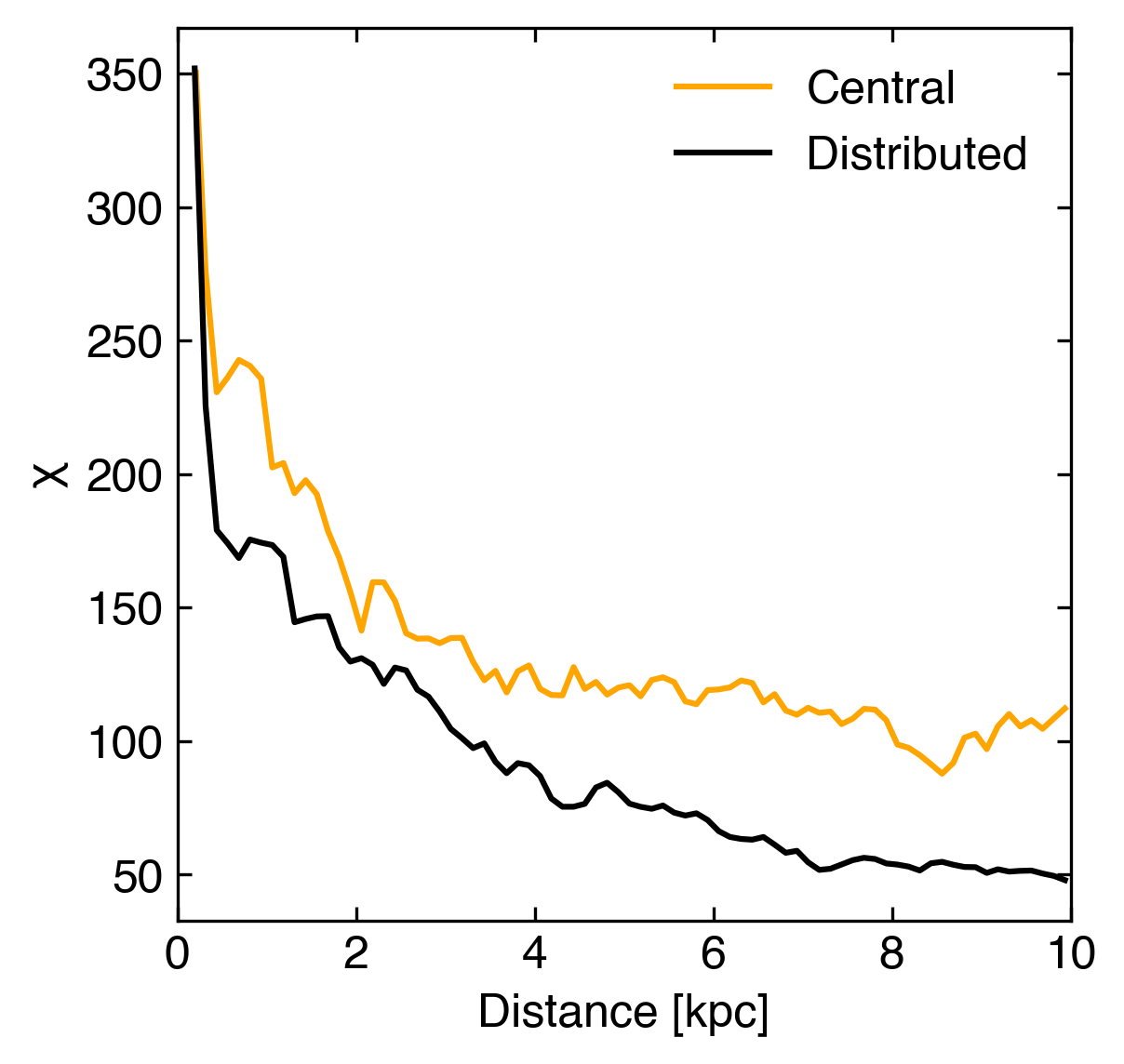}
    \caption{$\chi$, the density contrast between the cool and hot phases, as a function of distance for both feedback models.  Central feedback produces a hotter, more tenuous wind at most distances, reflected in the larger $\chi$ values.} 
    \label{fig:chi}
\end{figure}

\citet{gronke2018} framed the survival criterion as a competition between destructive dynamical mixing and cooling-induced mixing, which can fuel cloud growth. For the dynamical timescale, they adopted the ``cloud-crushing time" $t_{\rm cc} =\sqrt{\chi} (r_{\rm cl}/ v_{\rm w})$, where $\chi \equiv \rho_{cl}/\rho_w$ is the density contrast between the cloud and the hot wind, because clouds are destroyed over a few $t_{\rm cc}$ in the adiabatic limit \citep{Klein1994}. Fig. \ref{fig:chi} shows the value of $\chi$ measured in our simulations. The relevant cooling timescale has undergone some evolution. Initially, \citet{gronke2018} suggested the cooling time of the mixed gas, evaluated at the geometric mean of the hot wind and cool cloud temperatures. \citet{farber2022} suggested a refined estimate, $t_{\rm cool,minmix} \equiv t_{cool}(\sqrt{T_{\rm min, cool}\ T_{\rm w}}, p_0)$, where $T_{\rm min,cool}$ is the temperature between $T_{\rm cl}$ and $T_{\rm w}$ for which $t_{\rm cool}$ is minimized. Consequently, \citet{farber2022} predict cloud survival when $t_{\rm cool,minmix}< t_{\rm cc}$.
This can be rearranged into a minimum size criterion \citep{farber2022}
\begin{equation}
r_{\rm crit, cc} = \frac{v_{\rm w} t_{\rm cool,minmix}}{\chi^{1/2}}.
\end{equation}

\begin{figure*}
    \centering
    \includegraphics[width=0.49\linewidth]{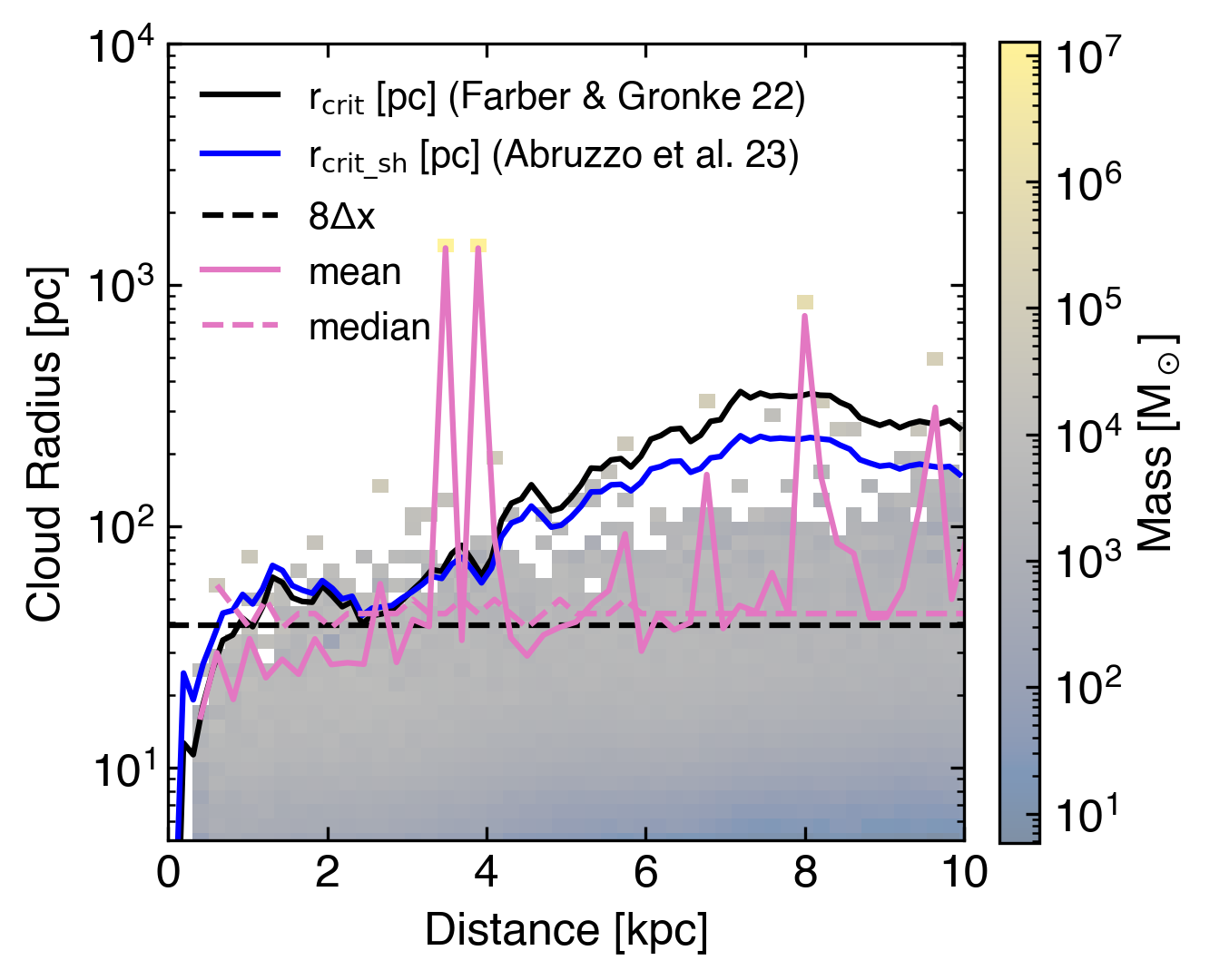}
    \includegraphics[width=0.49\linewidth]{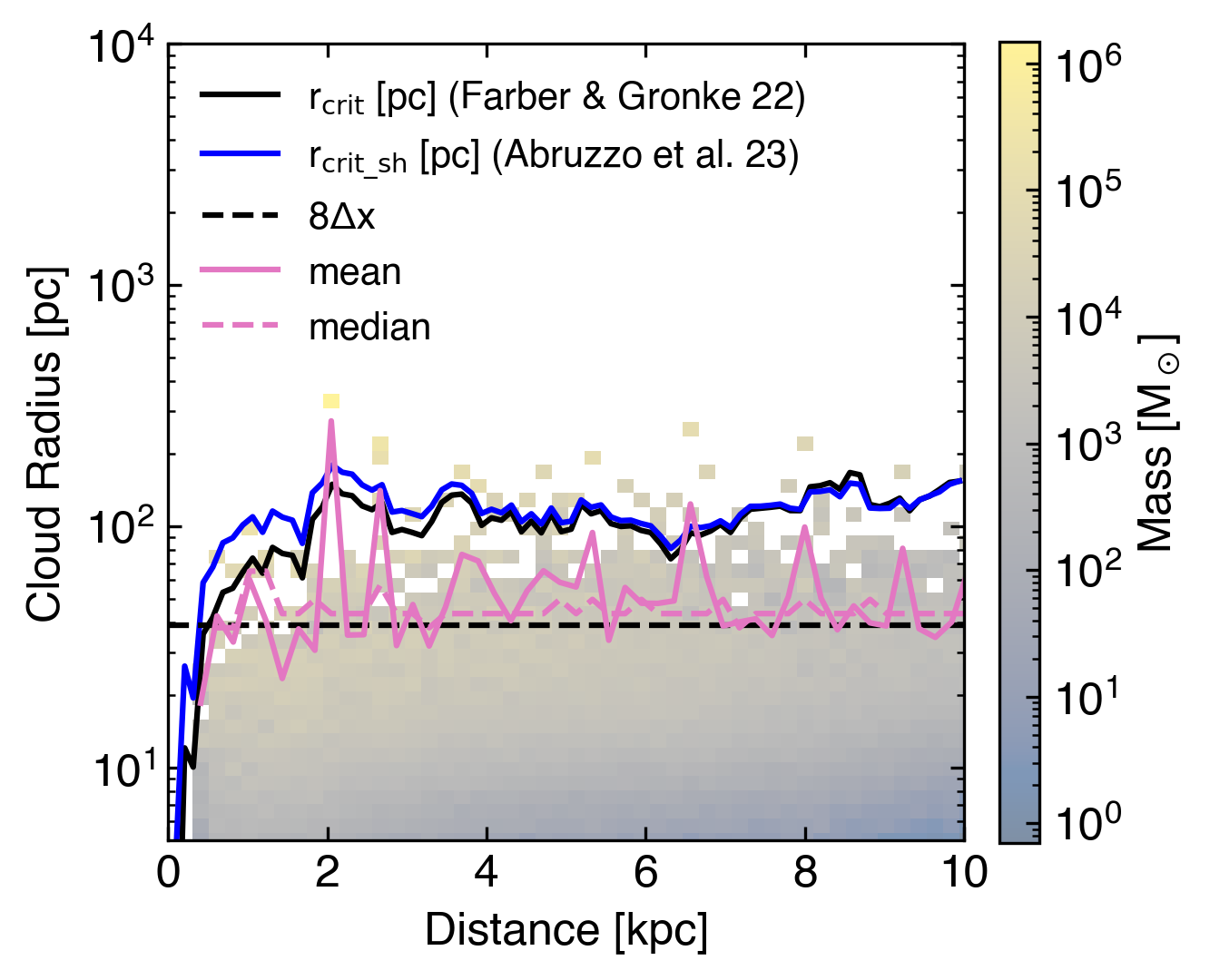}
    \caption{2-Dimensional mass-weighted histogram of cloud radii and distance (left: distributed feedback; right: central feedback).  The critical radius criterion from \citet{farber2022} is shown in black, and that from \citet{abruzzo2023b} in blue ($\alpha = 10$).  Cloud median radii are for resolved clouds only.  The limit for a ``resolved" cloud is shown as a dashed horizontal line at $40\,\mathrm{pc}$.}
    \label{fig:rcrit_2048}
\end{figure*}

More recently, \citet{abruzzo2023b} proposed an alternative criterion based on the observation that mixing drives either cloud growth or destruction. A cloud only grows (and survives), when it absorbs material from the wind.
For wind material that mixes with the cloud to be absorbed, the material must cool to ${\approx}T_{\rm cl}$ before it advects past the end of the cloud. In other words, $t_{\rm cool,minmix}$ must be shorter than this advection timescale, which is related to $t_{\rm shear} = r_{\rm cl}/v_{\rm w}$.
Thus, clouds should survive when $t_{\rm cool,minmix} < \alpha t_{\rm shear}$, where $\alpha$ empirically accounts for elongation of the cloud and decreases in the material's relative speed from mixing and cloud acceleration.
The corresponding size criterion is 
\begin{equation}
r_{\rm crit,sh} = v_{\rm w}\, t_{\rm cool,minmix}\, \alpha^{-1}.
\end{equation}
This criterion makes predictions that are more consistent with other works \citep[e.g.][]{li2020, sparre2020} for $\chi > 10^2$.
We adopt $\alpha\approx 10$ so that $r_{\rm crit,sh}=r_{\rm crit,cc}$ for $\chi=10^2$.

Unlike the idealized simulations used to derive both survival criteria, our clouds are not in pressure equilibrium with the wind \citep[see][Figure 6]{Schneider2024}. Thus, in order to make a comparison between the clouds in our simulations and the relevant criteria, we must make certain assumptions in order to determine $r_\mathrm{crit}$. For our purposes, we choose to use the geometric mean of the hot and cool phase pressure, and the shear velocity between the hot and cool phase, as calculated using the density-weighted median values shown in \cite{Schneider2020, Schneider2024}.

Figure \ref{fig:rcrit_2048} plots these criteria on 2D cloud radius--distance histograms, along with the minimum cloud resolution requirement, $8\Delta x$.  Since most clouds in either simulation do not meet the survival criteria, we expect that the majority of the cloud population will not grow as it moves out.  At very small distances, this may be an artifact of the limited resolution rather than caused by the underlying physics, i.e. the minimum $r_\mathrm{crit}$ for survival (as measured by either criterion) is below the resolution threshold for a resolved cloud. However, at larger distances as the value of $r_\mathrm{crit}$ increases, we can evaluate the implications of these criteria.

First, we note that at small distances ($< 3-4\kpc$), the critical radius for cloud survival is significantly larger in the central feedback simulation than in the distributed, particularly the value devised by \cite{abruzzo2023b}. As a result, although resolved cloud sizes for both simulations are similar in this region, we might expect significantly more cloud survival in the distributed feedback case (i.e. this could explain the trend seen in the left panel of \autoref{fig:numbers}). The scenario reverses at larger radii - while in the central feedback case the criterion stays roughly flat at around $100 - 200\pc$, in the distributed feedback case it continues to grow. This may explain why large clouds in the distributed feedback case do not gain mass as they move outward (i.e. the right panel of \autoref{fig:numbers}).

In this snapshot, there are a total of $1254$ resolved clouds in CGOLS IV, with a combined mass $3.76\times 10^{6} \Msun$ ($65\%$ of the total).  Of these resolved clouds $95$, with $1.63\times 10^{6} \Msun$, have radii greater than both critical radius criteria.  In CGOLS V there are $3817$ resolved clouds with a total mass $2.62\times 10^7\Msun$ ($94\%$ of the total). A total of $544$ of these, with $2.51\times 10^7\Msun$, have radii greater than both criteria.  By both these criteria $96\%$ of the resolved distributed feedback cloud mass resides in clouds that should be growing, as does $43\%$ of the resolved central feedback cloud mass. Thus, we expect to see the net mass flux in cool clouds for the distributed simulation increasing (or at least surviving), while for the central feedback case we expect to see net cloud destruction. Again, these expectations are consistent with the result shown in the right panel of \autoref{fig:numbers} that the mass flux in CGOLS V is flat with radial distance, whereas it slightly decreases in the CGOLS IV case.

In addition to examining the validity of the cloud survival criteria on the entire population of clouds, we can also focus in more detail on a single cloud that meets both survival criteria, in order to explore the extent to which predictions based on idealized simulations can provide insight to its evolution (or fail to do so). One example of a cloud whose radius clearly exceeds both critical values is the $10^{5.5}\Msun$ cloud located at $\sim 8 \kpc$ in the distributed simulation discussed at the end of the previous section.  As already hinted at in Section \ref{subsec:time_evolution}, the measured total velocity of this cloud as determined in the catalog differs substantially from the actual speed of the object whose center-of-mass is being tracked. For example, during the $26$\textendash$34\Myr$ interval that the cloud can be tracked, its average radial speed (as measured in the cloud catalog) is $546\kms$ with a standard deviation of only $13\kms$.  On the other hand, if we calculate the average radial speed from the distance traveled by the cloud between each snapshot, i.e. in \autoref{fig:2dtime_mean}, we find $308\kms$, with a much larger standard deviation of $195\kms$.  Clearly, the object we are calling a cloud in this case is not the sum of its parts. In fact, this discrepancy likely points to the way that cool mass is being entrained -- while the body of the cloud is moving out relatively slowly, cool mass accretion from the  hot phase is occurring stochastically along the cloud's tail, a result commonly seen in idealized cloud simulations. As we will demonstrate in the following analysis, much of this added mass must not remain part of the cloud; rather the tail is constantly fragmenting and breaking away. In this sense, the cloud we identify is more like a density wave, in which the pattern speed need not match the speed of the gas making it up.

Given its properties, we can roughly estimate the expected mass growth for this cloud based on idealized simulations of sheer mixing layers and cloud-wind setups. From dimensional analysis, the equation for mass growth takes the form 
\begin{equation} 
\dot{m} \sim v_{mix}A_{cl} \rho_{hot},
\label{eq:massgrowth}
\end{equation} 
where $A_{cl}$ is the cloud effective area and $v_{mix}$ the mixing velocity by which hot gas with density $\rho_{hot}$ is advected onto the cloud \citep{gronke2020a}. Note that the turbulent cloud-wind interaction for a cloud with volume $V$, is expected to yield an effective area $A_{cl} \propto V^{D/3}$, with fractal dimension $D>2$ \citep{fielding2020, gronke2020a}, though for simplicity in the following analysis we will simply use D = 2 as a lower limit.  The mixing velocity can be estimated from the relationship 
\begin{equation} 
v_{mix} \propto v_{turb}^{3/4}\left(\frac{L}{t_{cool}}\right)^{1/4},
\end{equation}
where $L$ is the integral scale of turbulence, $v_{turb}$ is the turbulent speed, $t_{cool}$ is the cooling time, and all quantities are evaluated for the cloud \citep{tan2021}. Substituting the derived cloud radius for $L$, $724\pc$, and estimating $v_\mathrm{turb}$ from the shear velocity between the cloud and wind, $v_\mathrm{turb} \sim f v_\mathrm{sh} \approx 0.1 (v_\mathrm{hot} - v_\mathrm{cl}) = 25\kms$, we obtain $v_{mix} \approx 15\kms$ for the cloud \citep[see also][]{tan2023}. We note that the eddy turnover time ($L / v_\mathrm{turb}$) is $\sim 29\Myr$, which is far longer than the $\sim 0.006\Myr$ cooling time \citep[justifying the use of $t_{cool}^{-1/4}$ in the $v_{mix}$ calculation][]{tan2021}.  \footnote{We also note that although $v_{mix}$ saturates with $\mathcal{M}$ \citep{yang2023}, at our cloud $\mathcal{M} \sim 1.5$ the actual value will differ by a factor of order unity.} With this value for $v_{mix}$, along with the effective area $6.6 \kpc^2$ and the median hot gas density of $5\times10^3\Msun\,\kpc^{3}$, we predict an $\dot{m} \sim 500\Msun\,\mathrm{Myr}^{-1}$. In fact, from 26 - 27 Myr we find growth of $\sim 2500\Msun$, while between 30 - 31 Myr the mass decreases by $5\times10^4 \Msun$. 

The fact that the changes in cloud mass between snapshots differ by an order of magnitude from our estimates based on idealized calculations hints at the fact that large scale fragmentation and coagulation events are likely more important in determining overall cloud growth and survival in this regime. Indeed, other work has shown that although the mass flux onto clouds on small scales  appears to follow the relations above quite well, they do not predict the larger-scale behavior of clouds \citep{tan2023}. In this sense, focusing on this large cloud in our simulation may be more analogous to looking at the ``biggest clump" in a turbulent box simulation \citep[i.e.][figure 11]{gronke2022}. That study noted a certain amount of stochasticity in large clump growth, similar to what we find with this cloud.  They additionally found a mach number dependence to the idealized cloud crushing $r_{crit}$ expression that increases $r_{crit}$ with  $\mathcal{M}$, since turbulence  makes shattered droplets less likely to coagulate.  This contrasts with the mach number dependence found by other groups in wind tunnel setups \citep{li2020}, in which case increased $\mathcal{M}$ has a stabilizing effect on the cloud via compression.  Although in this work we do not track individual clouds, since the outflowing hot phase in these global simulations has a net radial outflow along with a turbulent component, we expect these two competing effects to decrease each other's importance.

\subsection{Comparison to Larger Scales} \label{subsec:HVC}

In recent years, several other authors have investigated the properties of clouds in a number of different simulations \citep{gronke2022, ramesh2023, ramesh2024, tan2023}. As highlighted in the introduction, the simulations we investigate in this paper lie at the intersection of higher resolution idealized studies and larger scale cosmological simulations. In previous sections we have primarily addressed the way our results align with the results of smaller-scale simulations. In this subsection we briefly highlight some additional similarities and differences between the trends we show here and those seen in larger-volume simulations.

In the Milky Way and other galaxies, outflowing cool gas clouds are thought to be a source of observed near circumgalactic medium (CGM) high velocity clouds (HVCs) \citep[e.g.][]{putmanGaseousGalaxyHalos2012}. Using cosmological zoom-in simulations of Milky Way-like galaxies from the GIBLE project, \citet{ramesh2024b} estimate that $40-60\%$ of cool gas clouds in the CGM of these galaxies arise from galaxy outflows. Although the simulations we investigate here are for lower-mass galaxies, we do expect these outflows to be a significant source of cool gas in the inner CGM based on the cloud velocities. By extrapolating from the cloud population near the simulation boundary at $10\kpc$, we expect to have between $650$ and $5000$ clouds per Myr populating the CGM with distributed feedback, and between $150$ and $1000$ clouds per Myr with central feedback. These estimates are for clouds above our resolution limit (8 cell minimum cloud radius), and the upper end of the range accounts for the limited solid angle coverage of our data. Many of the larger clouds in these populations should survive well into the CGM.

\citet{ramesh2023} used the TNG50-1 cosmological simulation to look at cold gas clouds in 132 Milky Way-like galaxy CGMs.  With an average baryon mass resolution element of $8\times 10^4\, \mathrm{M_\odot}$, the cloud population studied in these simulations is primarily much larger than what we model. They found on order $10^2$ fully ($10^3$ marginally) resolved clouds in the typical CGM (with any cloud made up of 10 or more Voronoi cells meeting their fully-resolved threshold).  While these numbers are insensitive to galaxy stellar mass (ours is $0.5$ dex lower than their minimum galaxy mass), they are sensitive to the specific star formation rate (sSFR), with higher sSFR galaxies harboring greater numbers of clouds (our sSFR, $2\times10^{-9}\,\rm{yr^{-1}}$ is 1.1 dex higher than their maximum sSFR $\sim 1.6\times10^{-10}\,\rm{yr^{-1}}$).  

%Compounding the difficulty in comparing results, the TNG50-1 simulation uses a simplified stellar feedback driven wind model, which limits the modeling of inner CGM clouds, while our simulation does not model AGN feedback, the presence of magnetic fields, or cloud contributions from satellite galaxies.

Although they find that the typical CGM cloud has a mass of $\sim 10^6\,\Msun$, this value is closer to $\sim 10^5\,\Msun$, when including marginally resolved clouds, and decreases with increasing resolution, similar to the trend we describe in \autoref{sec:resolution}.  Tellingly, these lower mass clouds tend to be found at smaller galactocentric distances, which is also consistent with the inverse dependence of $r_{crit}$ on the pressure: a higher pressure environment, such as that in the outflow simulations we have investigated, will produce a smaller-sized cloud population than in the CGM. The mass PDF, for fully resolved clouds, has a power-law shape above $\sim 1.6\times10^6\Msun$ with a slope $\sim -1.5$ \citep[see Figure 4 of][]{ramesh2023}.  This is shallower than the $-2$ slope of our mass PDF, and may reflect a different population origin for the largest CGM clouds. However, including marginally resolved clouds steepens the slope above the distribution mode. Using the GIBLE simulations, with a significantly lower average CGM resolution element of $\sim 2\times 10^3 \Msun$, \citet{ramesh2024} find that the number of clouds as a function of mass $M$ has a power law slope of shape $\rm{log}_{10}[N(M)] \propto M^{-1}$, which is consistent with what we find in this work.

Much like the large cloud we discussed at the end of \ref{subsec:R_crit}, GIBLE CGM clouds show non-monotonic mass evolution, with fragmentation and merging playing important roles. \cite{ramesh2024b} find rates of $\sim 0.02 \Myr^{-1}$ ($\sim 0.03\Myr^{-1}$) for mergers (fragmentations) occurring in $\sim 10^5\Msun$ clouds, which is much lower than what we find ($\sim 0.5\Myr^{-1}$ fragmentation events for the large cloud). However, because of the time resolution of the snapshots, both our estimated rates and the rates found in GIBLE are underestimates, and the parameter space we probe is very different from that in the CGM. For example, the pressure imbalance between cool and hot phases in our study is more extreme than what GIBLE probes in the CGM ($\Delta \rm{log}_{10}(\rm{P [K cm^{-3}]}) \sim -0.5$ versus a minimum value of $\sim -0.24$ in GIBLE), and the velocity shear between the hot and cool gas is much higher in outflows. These differences are likely to impact merger/fragmentation rates and make direct comparisons difficult.

\section{Conclusions}\label{sec:conclusion}

In this work we have described population attributes of cool gas clouds in outflows generated by global galaxy simulations from the CGOLS suite. In particular, we have investigated the cloud properties from two simulations that differ in their application of stellar feedback: one models a nuclear starburst (``central feedback"), while the other has clusters dispersed throughout the disk (``distributed feedback").

Although the cloud populations are qualitatively similar, we find that differences in feedback model do impact some cloud properties. In particular, central feedback results in fewer and denser clouds than in the population generated by distributed feedback (Section \ref{subsec:mass_function}). That said, both models produce a spectrum of clouds with properties that change in similar ways as they move out from the galaxy (Section \ref{subsec:radial_dep}).

In addition, we find that overall statistical measures of cloud properties are broadly consistent with theories of turbulence. For example, in \autoref{fig:mass_pdf} we demonstrate that in the limit where clouds are well-resolved, the cloud mass function follows a power law with a slope of $\sim-2$, consistent with turbulent fragmentation (and indicating a similar mass in clouds per logarithmic bin). We also show that the internal velocity dispersion of clouds is consistent with being set by turbulence, with the mean of the cloud populations following a version of Larson's Law (\autoref{fig:2d_sigma_R_cone}), and that the first-order velocity structure function estimated by the cloud-cloud velocity dispersion reflects the driving scale of turbulence (\autoref{fig:vel_struct_bins2}).

We also explore cloud properties in the context of recent theoretical criteria for $R_{crit}$ -- a minimum cloud radius for survival in winds -- and find that these criteria are good predictors of cool gas evolution in outflows generated in realistic galaxy environments, although their applicability to any particular structure identified as a cloud is less robust. As in turbulent-box simulations, we find evidence of cloud fragmentation and coagulation. We plan to explore this futher in future work by adding tracer particles to follow the time evolution of individual clouds, in order to allow further refinements to analytical $R_{crit}$ formulations in a global context.

Our approach can provide a ``missing link" between idealized wind tunnel simulations of cloud evolution and cosmological simulations of clouds in the CGM. Future work will track clouds further out into the near CGM, for better comparison with cosmological simulations, and to provide better subgrid recipes for winds.

\begin{acknowledgements}
E.E.S thanks Drummond Fielding, Max Gronke, Brent Tan, and other participants at the ``Modeling of Multiphase Media" conference for helpful discussions that improved the quality of this work. This research used resources of the Oak Ridge Leadership Computing Facility, which is a DOE Office of Science User Facility supported under Contract DE-AC05-00OR22725, using Titan allocation INCITE AST125 and Summit allocation CAAR CSC380. This research was supported in part by the University of Pittsburgh Center for Research Computing, RRID:SCR\_022735, through the resources provided. Specifically, this work used the H2P cluster, which is supported by NSF award number OAC-2117681. E.E.S. acknowledges support from NASA TCAN grant 80NSSC21K0271, NASA ATP grant 80NSSC22K0720, and the David and Lucile Packard Foundation (grant no. 2022-74680). S.A.M. acknowledges support from Pitt PACC.
\end{acknowledgements}

%\facilities{}

\software{\texttt{Cholla} \citep{Schneider15}, \texttt{numpy} \citep{VanDerWalt11}, \texttt{scipy} \citep{Virtanen2020},  \texttt{matplotlib} \citep{Hunter07},  \texttt{hdf5} \citep{hdf5}}

%\bibliography{windcloud}{}
%\bibliographystyle{aasjournal}

\appendix
\section{Resolution}\label{sec:resolution}

In this appendix, we explore the extent to which our results may be impacted by numerical resolution. Note that in this appendix all the cloud catalogs are generated using the full simulation volume.  Clouds within the biconical region perpendicular to the galaxy plane are then selected for the following analysis.

\begin{figure*}
    \centering
    \includegraphics[width=0.49\linewidth]{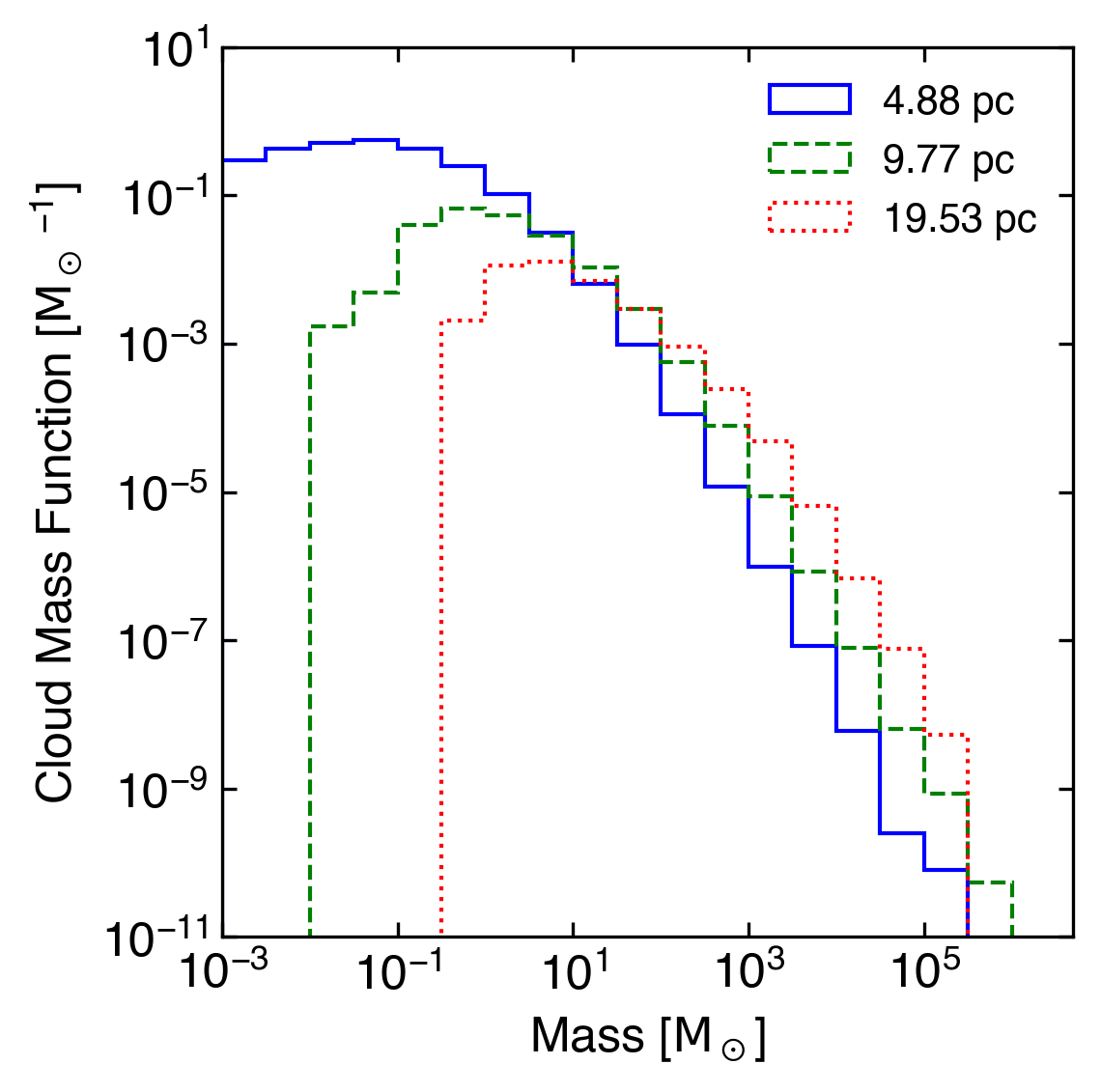}
    \includegraphics[width=0.49\linewidth]{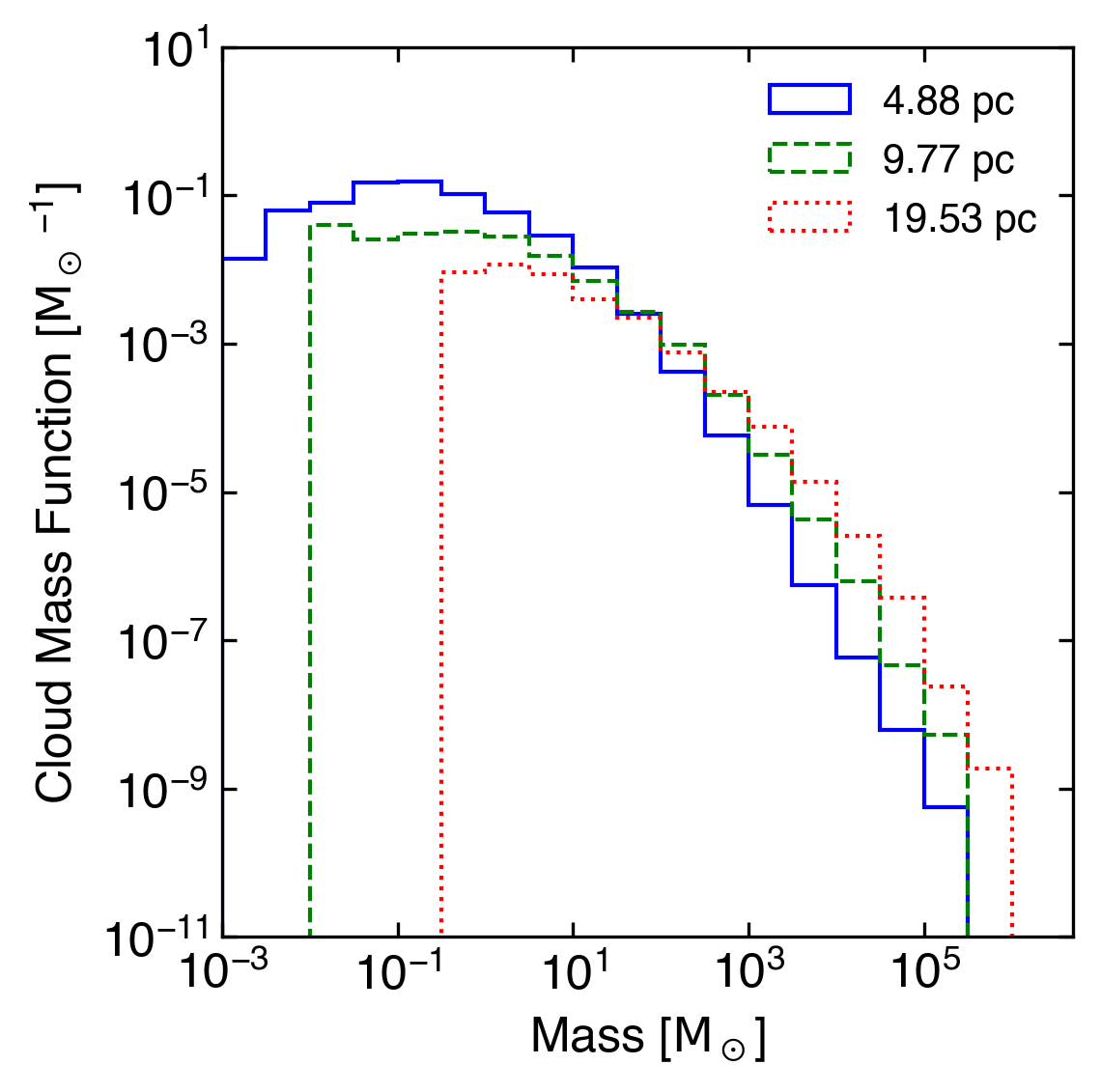}
    \caption{Resolution changes to the mass function for the case of distributed feedback in the left panel, and central feedback in the right panel. The low mass cutoffs seen for the two lower resolution simulations reflect the volume of single cells.} 
    \label{fig:mass_func_resolution}
\end{figure*}

As noted in Section \ref{subsec:mass_function}, the slope of the mass function of clouds turns over from a $-2$ power-law at low masses. Although it is tempting to identify this turnover as a result of smaller clouds being more susceptible to destruction, we do not have sufficient numerical resolution to make that case robustly. In Figure \ref{fig:mass_func_resolution} we see that the mass function peaks at higher values of mass with decreasing resolution, while at the low mass end, the distribution is always truncated by the minimum grid cell volume. (The lowest distributed feedback density is $\rho \approx 10^{-5} \mathrm{M_\odot \pc^{3}}$, therefore $\rho (20 \pc)^3 \approx 10^{-1.1} \mathrm{M_\odot}$.) However, with both feedback models the power law relationship is independent of resolution when we account for the higher turnover mass with decreasing resolution.  In the case of distributed feedback, the fitted slope values are $-2.11$ at $\Delta x=19.53\pc$, $-2.04$ at $\Delta x=9.77\pc$, and $-2.11$ at $\Delta x=4.88\pc$.  Corresponding values for the central feedback case are $-2.13$ at $\Delta x=19.53\pc$, $-1.84$ at $\Delta x=9.77\pc$, and $-1.92$ at $\Delta x=4.88\pc$. Thus, we conclude that the power-law slope derived in this analysis is robust.

\begin{figure*}
    \centering
    \includegraphics[width=0.49\linewidth]{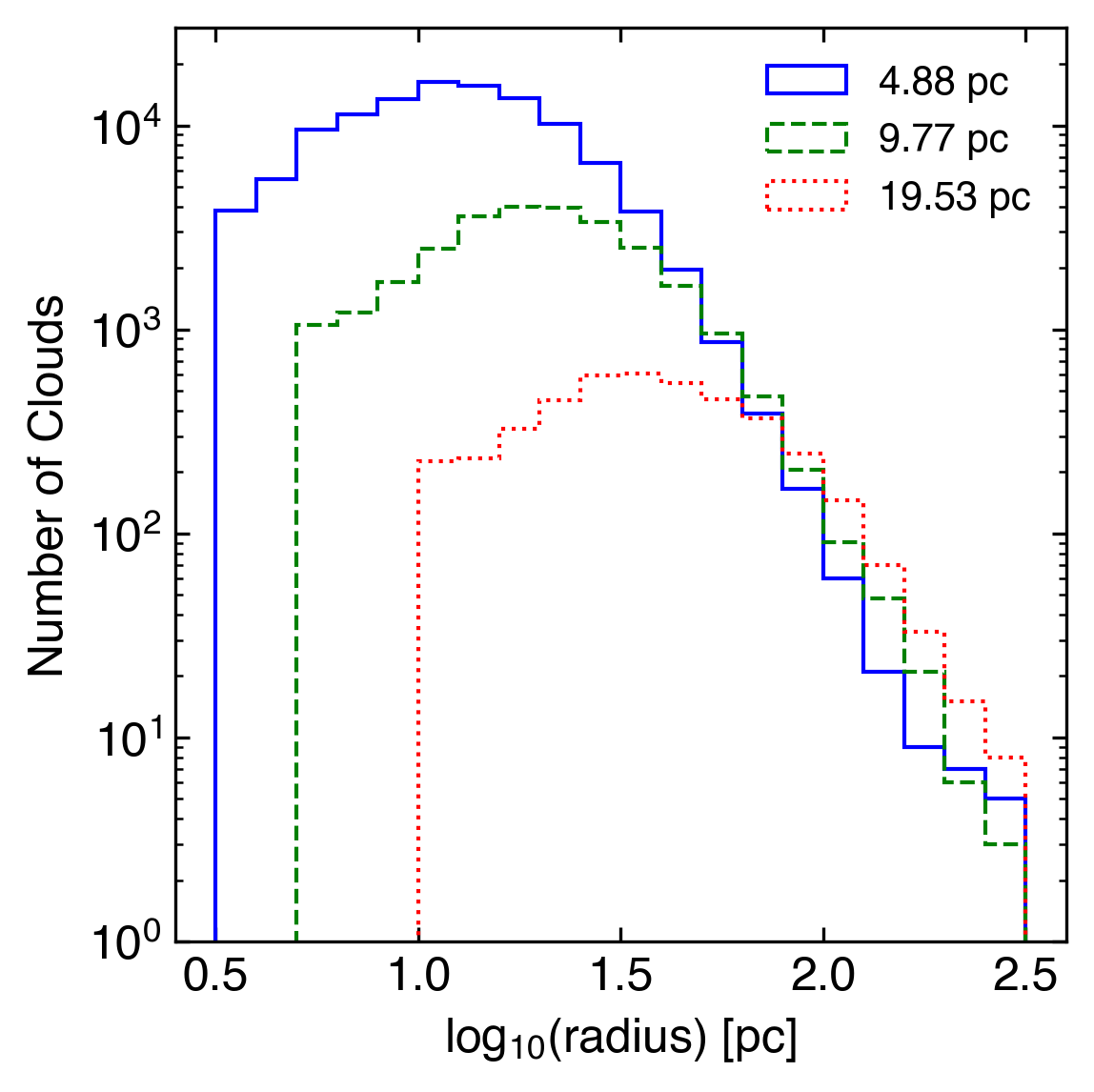}
    \includegraphics[width=0.49\linewidth]{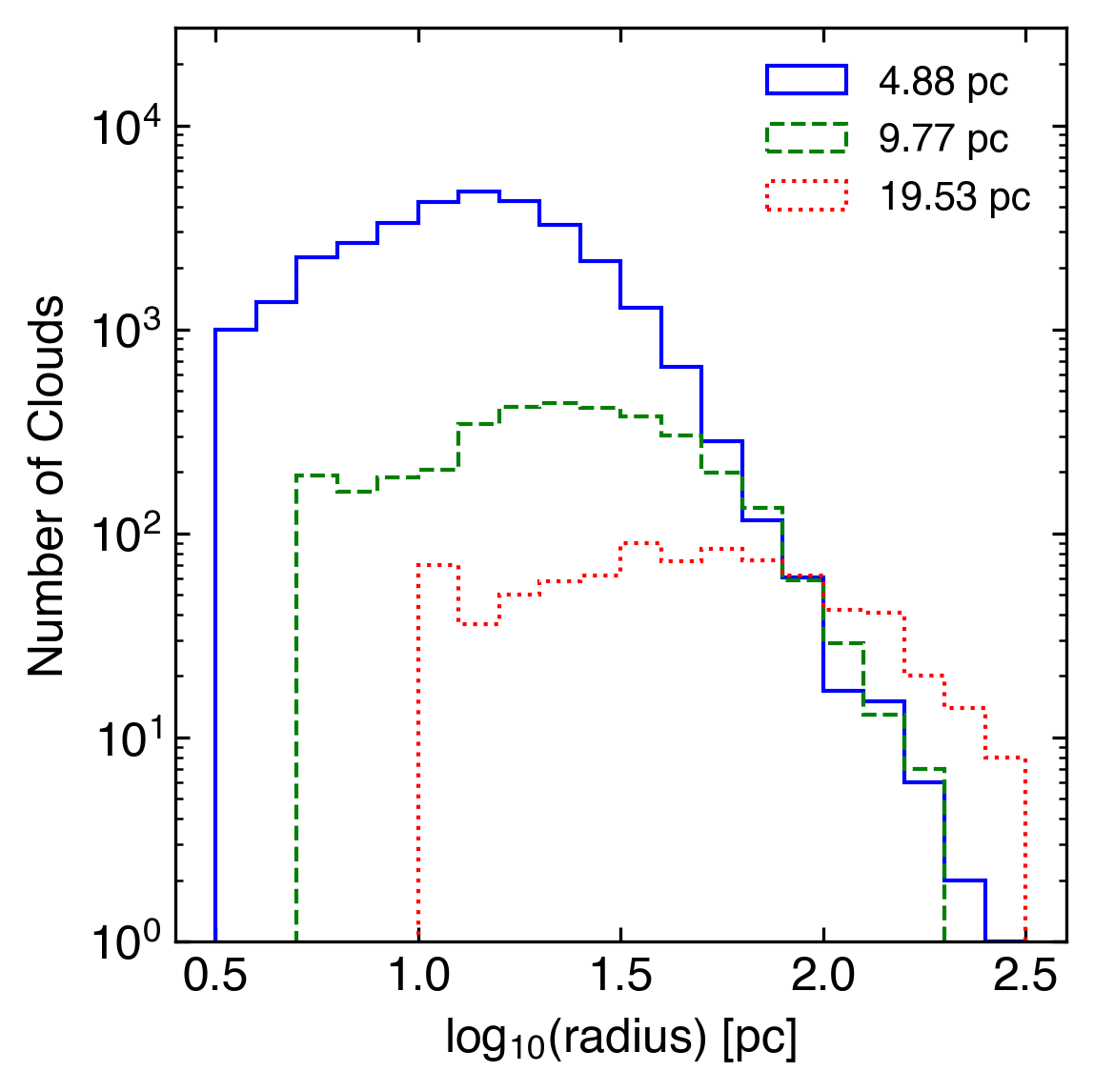}
    \caption{Resolution changes to the distribution of cloud radii for distributed feedback on the left panel, and central feedback on the right.} 
    \label{fig:rad_resolution}
\end{figure*}

Similarly, we see in \autoref{fig:rad_resolution} that the distribution of cloud radii is resolution-dependent, with the lowest resolution simulations generating preferentially larger clouds, and a clear trend towards many more smaller clouds as the grid resolution decreases. However, the radius histograms at the large end do appear to be converging between resolutions of 10pc and 5pc, with very similar numbers of clouds in the largest bins. In addition to these 1D histograms, we have also investigated numerical resolution effects in the 2D relationships described in \autoref{subsec:radial_dep}, and do not find substantial differences. In general, we do not find that the overall physical picture we describe changes with resolution.

\begin{figure*}
    \centering
    \includegraphics[width=0.49\linewidth]{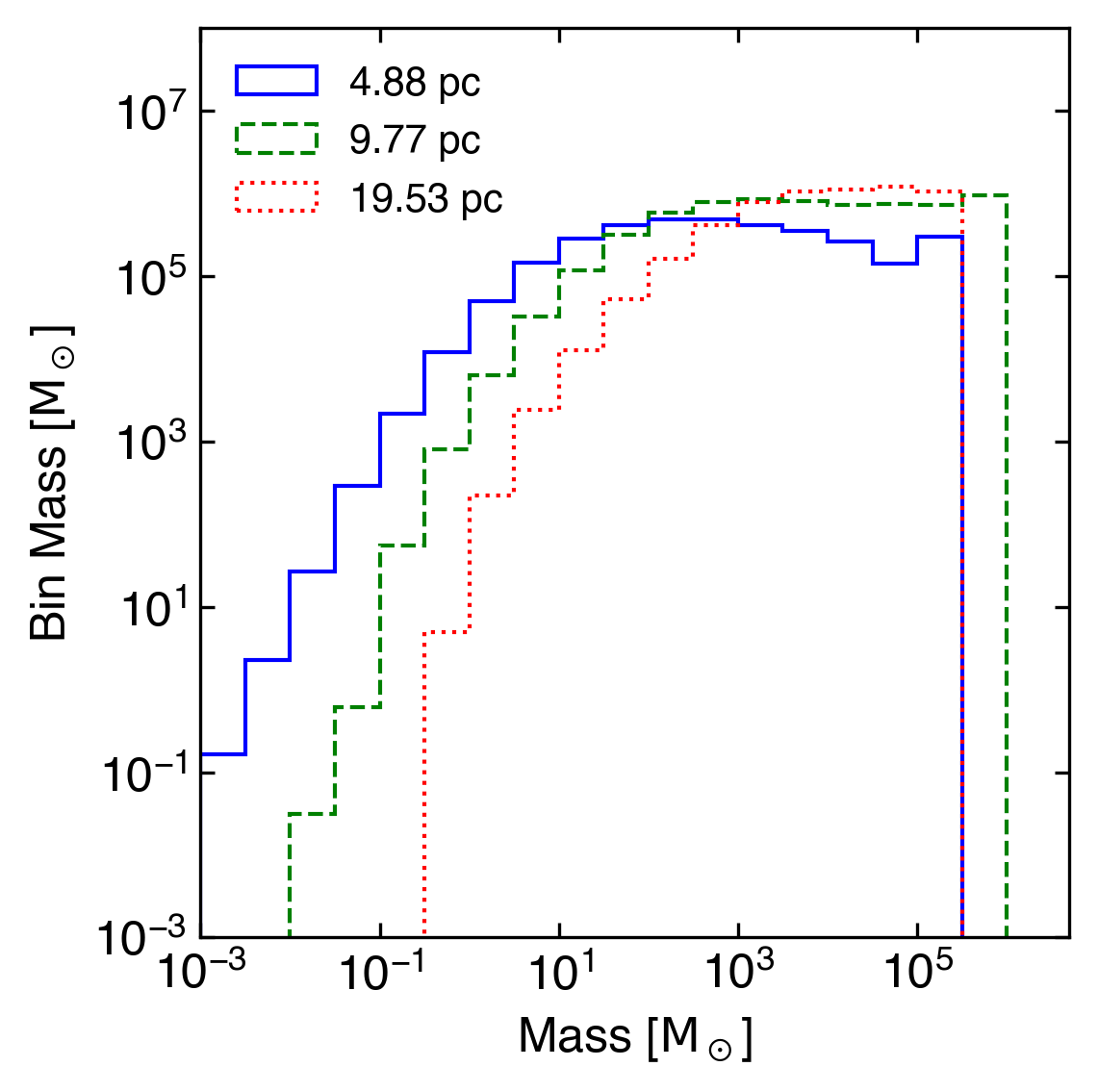}
    \includegraphics[width=0.49\linewidth]{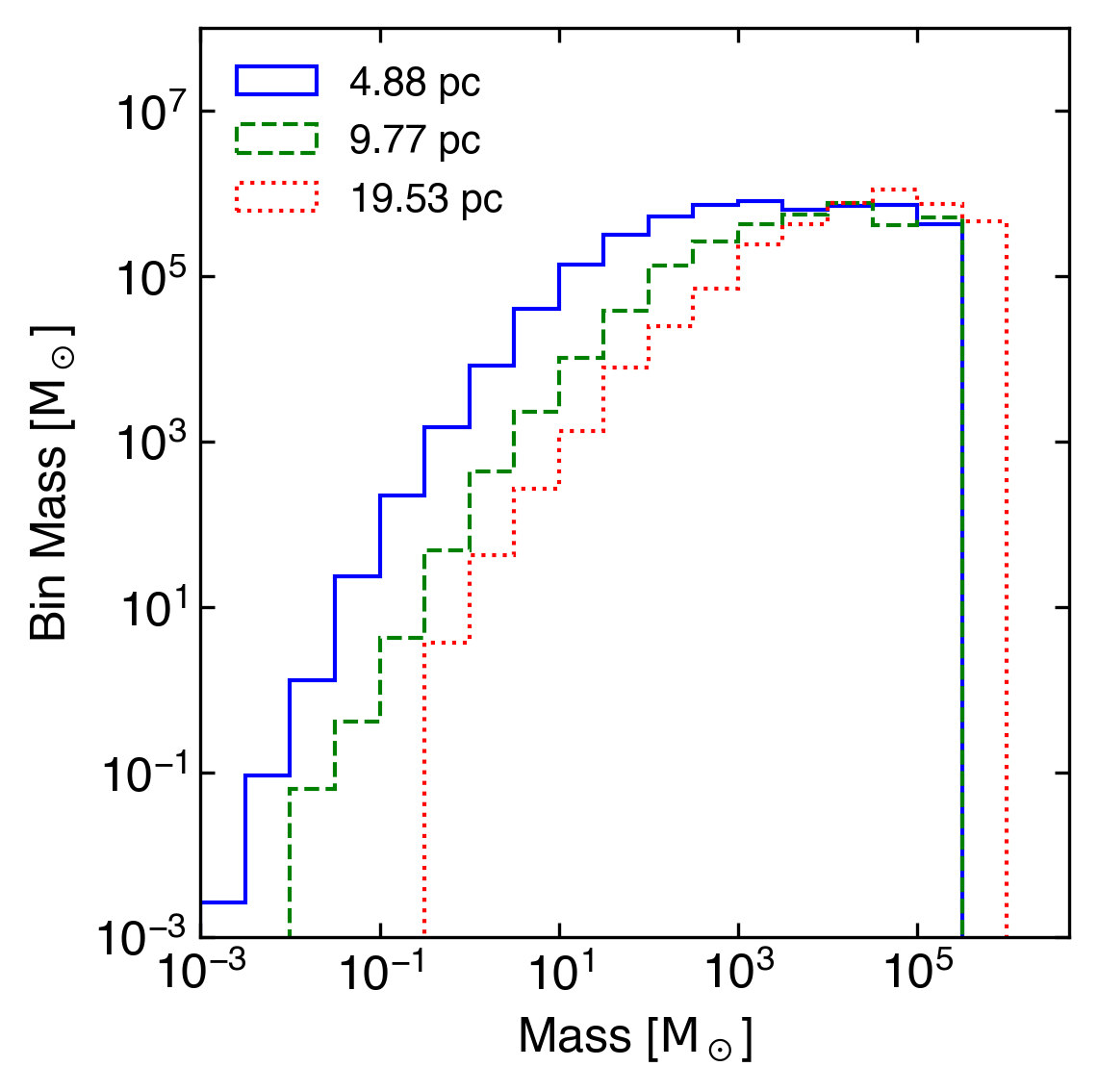}
    \caption{Resolution changes to the total mass in each bin for distributed feedback on the left panel, and central feedback on the right.  While some of the highest mass clouds are seen in the lower resolution data, cool mass is lost with decreasing resolution.} 
    \label{fig:mass_resolution}
\end{figure*}

\begin{figure*}
    \centering
    \includegraphics[width=0.49\linewidth]{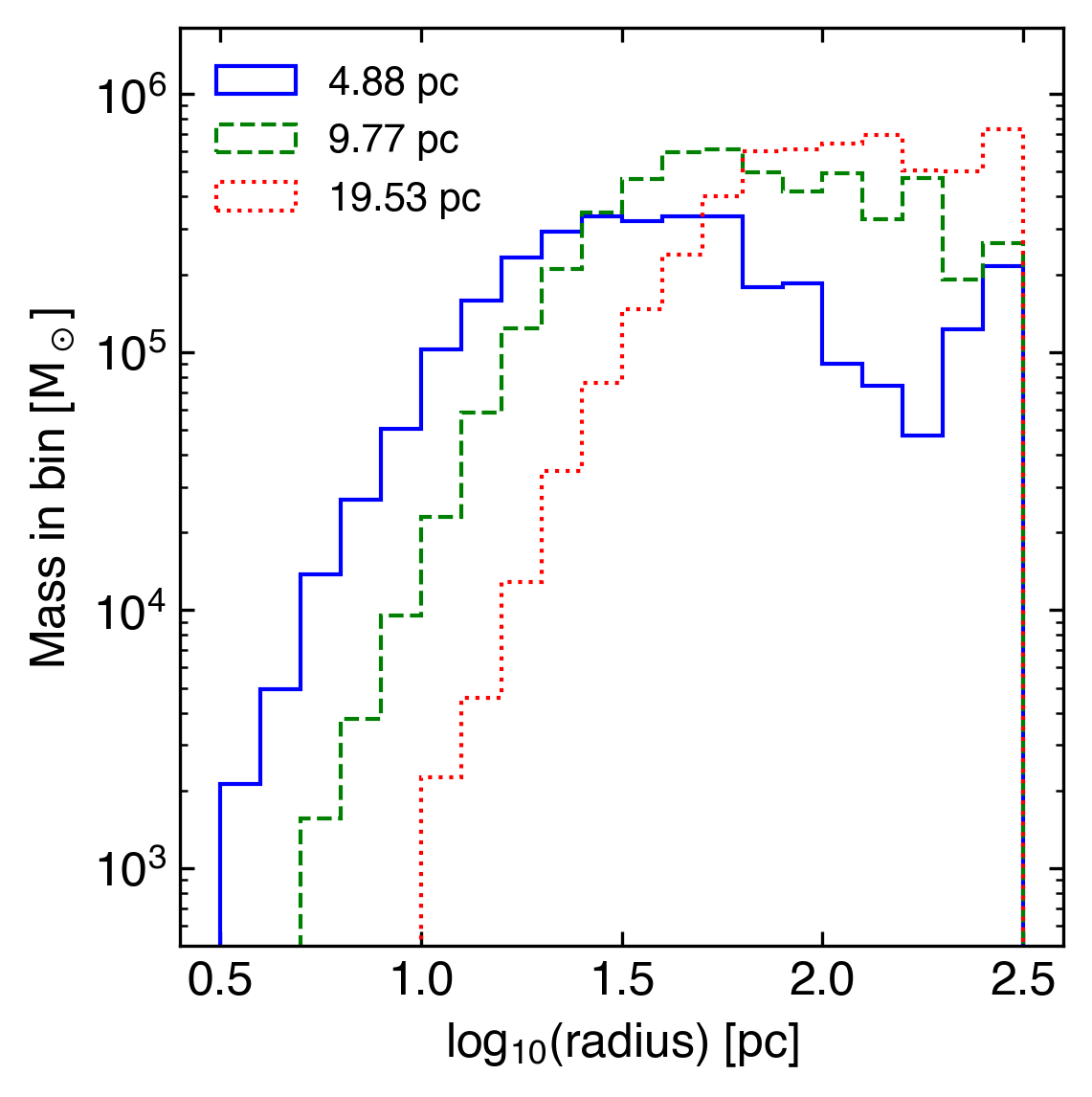}
    \includegraphics[width=0.49\linewidth]{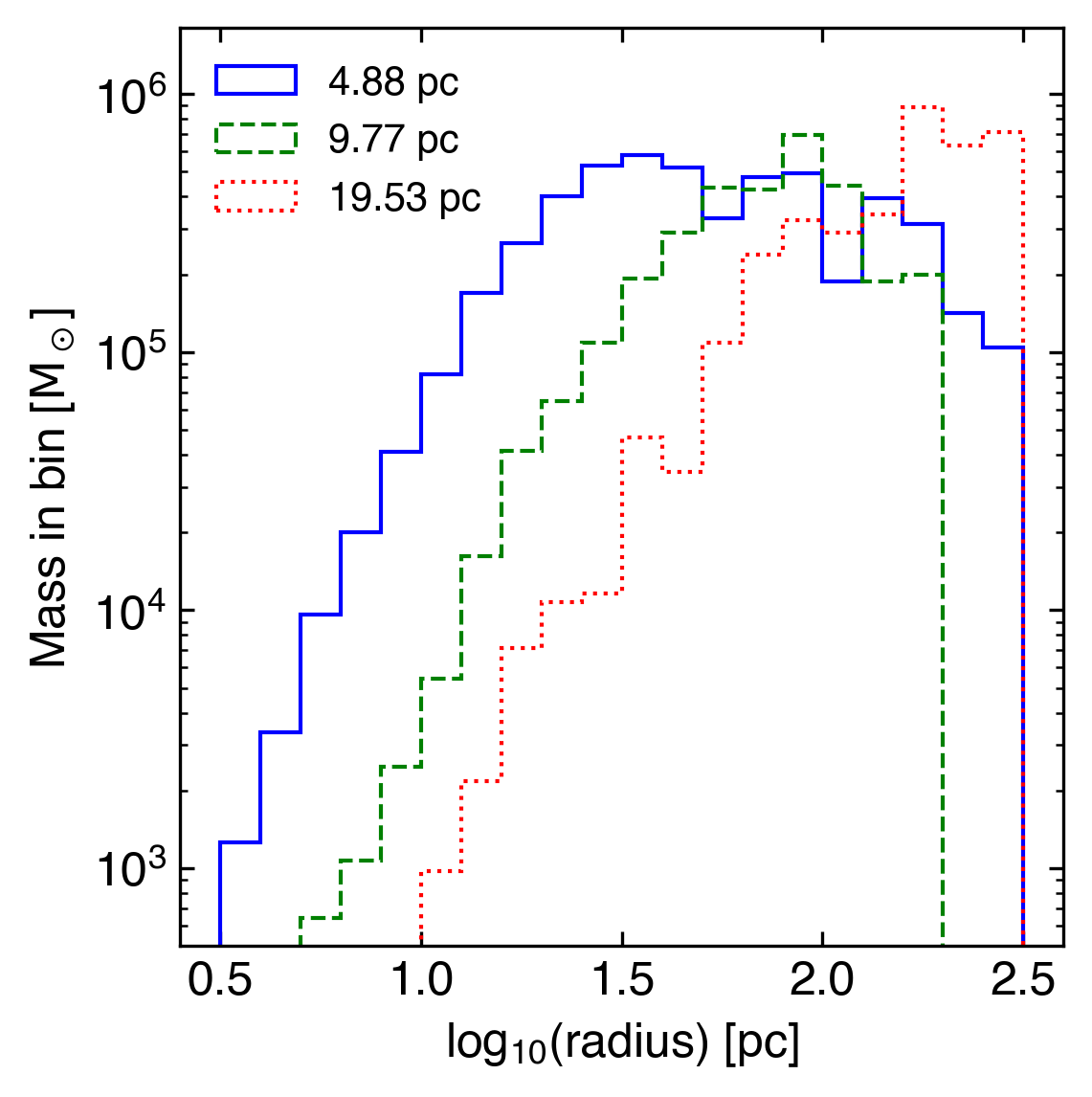}
    \caption{Resolution changes to how matter is distributed by cloud radius for distributed feedback on the left panel, and central feedback on the right.  The dip in $\sim 100 \rm{pc}$ clouds is less prominent (or disappears) at lower resolution.} 
    \label{fig:rad_mass_resolution}
\end{figure*}

On the other hand, an interesting trend emerges when looking at the mass-weighted versions of the mass and radius distributions. In Figure \ref{fig:mass_resolution}, we show the mass-weighted mass distribution, which again makes clear the trend of fewer low mass clouds in the under-resolved part of the distribution, and more mass in the high mass end. Upon careful inspection, we note that a dip appears in the 10pc resolution distributed feedback simulation around a mass of $10^{4.5}\Msun$, which becomes more prominent in the 5pc resolution simulation. A similar dip can be seen in the 5pc central feedback simulation, although it appears at slightly higher masses. Although the distributions in \autoref{fig:rad_mass_resolution} are less smooth, a similar dip appears in the highest resolution simulations at radii around $10^{2.2}\pc$ -- a similar scale to the $r_\mathrm{crit}$ criteria explored in \autoref{fig:rcrit_2048} for much of the simulation volume. This may indicate that at the highest resolutions, we are beginning to resolve a genuine break, where clouds below the destruction threshold are being destroyed, and those just above it are preferentially surviving and growing. We leave a further investigation of this trend with higher resolution simulations to future work.
    
\end{document}